\documentclass[%
reprint,
twocolumn,
groupedaddress,
noshowpacs,
longbibliography,
 amsmath,amssymb,
 aip
]{revtex4-1}

\usepackage[utf8]{inputenc}
\usepackage[T1]{fontenc}
\usepackage{mathptmx}
\DeclareMathAlphabet{\mathcal}{OMS}{cmsy}{m}{n}

\usepackage{graphicx,dcolumn,bm}
\usepackage{multirow,booktabs,mathrsfs,amsthm}
\usepackage[caption=false]{subfig}
\newtheorem*{problem}{Learning problem}

\begin{document}

\title{Learning the Tangent Space of Dynamical Instabilities from Data}

\author{Antoine Blanchard}
\email[E-mail for correspondence: ]{ablancha@mit.edu}
\author{Themistoklis P. Sapsis}
\email[E-mail for correspondence: ]{sapsis@mit.edu}
\affiliation{Department of Mechanical Engineering, Massachusetts Institute of Technology, \\ Cambridge, MA 02139}%

\begin{abstract}
For a large class of dynamical systems, the optimally time-dependent (OTD) modes, a set of deformable orthonormal tangent vectors that track directions of instabilities along any trajectory, are known to depend ``pointwise'' on the state of the system on the attractor, and not on the history of the trajectory.  We leverage the power of neural networks to learn this ``pointwise'' mapping from phase space to OTD space directly from data.  The result of the learning process is a cartography of directions associated with strongest instabilities in phase space.  Implications for data-driven prediction and control of dynamical instabilities are discussed.
\end{abstract}
\maketitle

\begin{quotation}
The optimally time-dependent (OTD) modes are a set of deformable orthonormal tangent vectors that track directions of instabilities along any trajectory of a dynamical system.  Traditionally, these modes are computed by a time-marching approach which involves solving multiple initial-boundary-value problems concurrently with the state equations.  However, for a large class of dynamical systems, the OTD modes are known to depend ``pointwise'' on the state of the system on the attractor, and not on the history of the trajectory.  We propose a neural-network algorithm to learn this ``pointwise'' mapping from phase space to OTD space directly from data.  The neural network produces a cartography of directions of strongest instability in phase space, as well as estimates for the leading Lyapunov exponents.  
\end{quotation}

\section{Introduction}
\label{sec:1}

The theory of dynamical systems has a long history of trying to elucidate one of the most important concepts in science and engineering: instability.  Dynamical instabilities are important because they can give rise to a variety of phenomena with unexpected, and even disastrous, consequences.  They occur in fluid mechanics \cite{chomaz2005global,schmid2007nonmodal}, climate dynamics \cite{penland1995optimal}, optics \cite{akhmediev2013recent}, and thermoacoustics \cite{balasubramanian2008thermoacoustic}, and come in many shapes and forms.  Perhaps the simplest of all are instabilities arising from a linear mechanism, whose investigation traditionally involves linearizing the governing equations around a fixed point, and looking for unstable eigenvalues of the linearized problem \cite{guckenheimer2013nonlinear}.  Generalization of this concept to periodic orbits led to the well-known Floquet theory, in which stability of periodic trajectories is ascertained by computing the spectrum of the monodromy matrix \cite{guckenheimer2013nonlinear}.  The realization that episodes of transient growth cannot be predicted by linear or Floquet analyses came much later, and in turn gave rise to the theory of non-normal and transient instabilities \cite{trefethen1993hydrodynamic}.  This culminated with the introduction of the Lyapunov exponents and Lyapunov vectors, which provide a rigorous description of instability mechanisms in chaotic systems \cite{eckmann1985ergodic, legras1996guide}.  Since then, considerable effort has been devoted to the development of algorithms for computation of Lyapunov exponents and Lyapunov vectors from data.  Great strides have been made by the likes of Eckmann et al. \cite{eckmann1986liapunov}, Sano \& Sawada \cite{sano1985measurement}, Rosenstein et al. \cite{rosenstein1993practical}, Kantz \cite{kantz1994robust}, and Wolfe et al. \cite{wolfe2007efficient}, who proposed new methods to compute Lyapunov exponents and Lyapunov vectors from experimental measurements, or improved on existing ones.  

The key feature shared by these algorithms is that they monitor the fate of perturbations around a reference orbit for sufficiently long times, sometimes resorting to orthornormalization in order to prevent blow-up of the perturbation vectors.  But tracking the evolution of perturbations along a trajectory is nothing more than a time-marching approach in disguise.  More importantly, it does not utilize the fact that for a large class of dynamical systems, the $i$th Lyapunov vector $u_i$ depends pointwise on the state of the system (with the mapping $x \longmapsto u_i(x)$ being uniquely determined by the phase-space point $x$), and consequently the $i$th Lyapunov exponent---a quadratic form over $u_i(x)$---can be recovered by measure-averaging over many measurement points, rather than time-averaging over a long trajectory.  This was pointed out by Ershov and Potapov \cite{ershov1998concept}, who recognized the benefits of using measure-averaging in lieu of time-averaging, as the former allows for the measurement points to be arranged in arbitrary order, and even for them to belong to different trajectories.  These results show that if one were able to compute the pointwise function $u_i(x)$ from data, then one would immediately have access to a complete picture of the directions of instabilities at any point in the phase space, as well as accurate estimates for the Lyapunov exponents.

As far as we know, the only attempt to compute the map $u_i(x)$ was made in Ref. \onlinecite{blanchard2019analytical}, where we used the theory of slow invariant manifolds by Fenichel \cite{fenichel1979geometric} to derive analytical expressions for $u_i(x)$ in situations exhibiting slow-fast dynamics.  However, application of this method is limited because a) the dimensionality of the system cannot be too large for the manifold analysis to be tractable; and b) the system must have a well-defined separation of time scales, short of which Fenichel's theory becomes moot.  Of course, the method being analytical, there is no data component to it.  This is precisely what we set out to rectify in the present paper, with the introduction of a machine-learning algorithm that infers the map $u_i(x)$ from a large collection of state measurements.  

Machine-learning (ML) algorithms have pervaded virtually every area of science and engineering because of two reasons.  First, the amount of data available from experiments, field measurements, or numerical simulations has reached unprecedented levels.  Second, ML algorithms have proven to be extremely versatile and powerful from the standpoint of extracting patterns and information from tremendously complex systems that would otherwise remain inaccessible.  Applications of ML to dynamical systems include dimensionality reduction and flow-feature extraction \cite{sirovich1987turbulence, sirovich1987low, amsallem2012nonlinear, kaiser2014cluster}, discovery of governing equations \cite{brunton2016discovering, long2017pde, raissi2018hidden}, design of optimal control strategies \cite{duriez2017machine}, turbulence closure modeling \cite{duraisamy2019turbulence,brunton2019machine}, integration of partial differential equations \cite{lagaris1998artificial,sirignano2018dgm,lu2019deepxde}, and forecasting of dynamical instabilities in chaotic systems \cite{silk2011designing, haugaard2017predicting}.  The recent emergence of deep learning has led to a push in the latter direction, with recurrent neural networks \cite{laptev2017time,wan2018data} and reservoir computing \cite{pathak2017using,vlachas2020recurrent} leading the way.  

The learning algorithm that we propose in this paper produces a cartography of directions associated with strongest instabilities in phase space, from which the leading Lyapunov exponents can be extracted.  These directions coincide with the backward Lyapunov vectors of Legras \& Vautard \cite{legras1996guide} and the optimally-time-dependent (OTD) modes of Babaee \& Sapsis \cite{babaee2016minimization}.  (The equivalence between the two was established  in Ref. \onlinecite{blanchard2019analytical}.)  The potential of the learning algorithm in problems related to prediction and control of transient instabilities and extreme events is considerable, because the proposed method is fully data-driven, has no restricting assumptions other than invertibility, autonomy, ergodicity, and measure-invariance of the underlying dynamical system, and only requires state measurements as inputs.

The paper is organized as follows.  We formulate the problem in \S\ref{sec:2}, introduce the learning algorithm in \S\ref{sec:3}, followed by results in \S\ref{sec:4}, a discussion in \S\ref{sec:5}, and a conclusion in \S\ref{sec:6}.

\section{Formulation of the problem}
\label{sec:2}

\subsection{Preliminaries}
\label{sec:21}

We consider the autonomous dynamical system
\begin{equation}
\dot{x} = F(x), \quad x(t_0) =  x_0,
\label{eq:21}
\end{equation}
where $x$ belongs to a compact Riemannian manifold  $\mathcal{X}$ endowed with the Borel $\sigma$-algebra and a measure $\mu$, the map $F \colon \mathcal{X} \longrightarrow \mathcal{X}$ is sufficiently smooth to ensure existence and uniqueness of solutions, and overdot denotes differentiation with respect to the time variable $t$.  We assume that the transformation 
\begin{align}
S_t \colon  \mathcal{X} 	&\longrightarrow \mathcal{X} \nonumber \\
				x_0 	&\longmapsto    x(t) = S_t(x_0),
\label{eq:22}
\end{align}
sometimes referred to as the ``flow map'', is invertible, measure-preserving, and ergodic.  Measure-invariance is important from the standpoint of defining time-averages of scalar functions.  (This is the well-known Birkhoff ergodic theorem \cite{lasota1985probabilistic}.)  Measure-invariance and ergodicity are important to guarantee that time-averages and measure-averages coincide:
\begin{equation}
\lim_{T \to \infty} \frac{1}{T} \int_{t_0}^T f(S_t(x_0)) \mathrm{d}t = \int_\mathcal{X} f(x) \mathrm{d}\mu(x), 
\label{eq:23}
\end{equation}
for all $f \in \mathscr{L}^2_\mu(\mathcal{X})$. In words, these assumptions ensure that trajectories asymptotically settle into an attractor $\mathcal{A} \subset \mathcal{X}$ (which may be steady, time-periodic, quasiperiodic, or chaotic), and remain on that attractor (i.e., there is no ``switching'' between attractors).  

Infinitesimal perturbations around a given trajectory obey the variational equation
\begin{equation}
\dot{v} = L(x;v), \quad v(t_0) =  v_0,
\label{eq:24}
\end{equation}
where $v$ belongs to the tangent space of the manifold $\mathcal{X}$ at point $x$, denoted by $\mathcal{T}_x\mathcal{X}$, and $L(x;v) \triangleq \mathrm{d}F(x ;v)$ is the G\^ateaux derivative of $F$ evaluated at $x$ along the direction $v$.  In principle, the variational equation could be used to track directions of instabilities around trajectories.  In practice, however, this is impossible, because any collection of perturbations $\{v_i\}_{i=1}^r$ evolved with (\ref{eq:24}) for a sufficiently long time would see the magnitude of its members grow or decay exponentially fast, and the angle between them rapidly vanish.  

To compute a set of meaningful directions (or ``modes'') from the variational equation, Babaee \& Sapsis \cite{babaee2016minimization} proposed to enforce orthonormality of the $v_i$'s at all times.  One way to achieve this is to continuously apply the Gram--Schmidt algorithm to the collection $\{v_i\}_{i=1}^r$, starting with $v_1$ and moving down.  Blanchard \& Sapsis \cite{blanchard2019analytical} showed that the resulting vectors obey
\begin{align}
\dot{u}_i &= L(x;u_i) - \langle u_i, L(x;u_i) \rangle u_i \nonumber \\
& \quad - \sum_{k=1}^{i-1} [ \langle u_i, L(x;u_k) \rangle + \langle u_k, L(x;u_i) \rangle] u_k
\label{eq:25}
\end{align}
for $i \in\{1, \dots, r\}$, where the angle brackets denote a suitable inner product on $\mathcal{T}_x\mathcal{X}$.  In the above, we recognize the variational equation (the left-hand side and the first term on the right-hand side), appended with terms enforcing the orthonormality constraint (the last two terms on the right-hand side).  We also note that the summation index goes to $i-1$ rather than $r$, so that (\ref{eq:25}) assumes a lower-triangular form.  (This reflects the very structure of the Gram--Schmidt algorithm.)  The $u_i$'s have been referred to as the ``OTD modes'', and the collection $\{u_i\}_{i=1}^r$ as the ``OTD subspace'' \cite{babaee2016minimization}.  The terms ``subspace'' and ``modes'' are appropriate because the collection $\{u_i\}_{i=1}^r$ forms a real vector space, for which the $u_i$'s are an orthonormal basis.

A key property of the OTD modes is that they and the $v_i$'s span the same subspace.  The first OTD mode aligns with the most unstable direction, just like $v_1$ does.  The second OTD mode is not free to align with the second-most unstable direction, because it must remain orthogonal to $u_1$.  But the subspace spanned by $u_1$ and $u_2$ coincides with the two-dimensional subspace that is most rapidly growing (this is also the subspace spanned by $v_1$ and $v_2$).  For hyperbolic fixed points, the OTD subspace aligns with the most unstable eigenspace of the associated linearized operator \cite{babaee2016minimization}.  For time-dependent trajectories, the OTD subspace aligns with the left eigenspace of the Cauchy--Green tensor, which characterizes transient instabilities \cite{babaee2017reduced}.  As a result, the $i$th Lyapunov exponent $\lambda_i$ can be recovered from the $i$th OTD mode:
\begin{equation}
\lambda_i = \lim_{T \to \infty} \frac{1}{T} \int_{t_0}^T \langle u_i(t), L(x(t);u_i(t)) \rangle \,\mathrm{d}t.
\label{eq:26}
\end{equation}
The fact that the OTD modes track directions of instabilities along any trajectory has been utilized on multiple occasions, including in the context of prediction of extreme events in chaotic systems \cite{farazmand2016dynamical}, and design of low-dimensional controllers for stabilization of unsteady flows \cite{blanchard2019control,blanchard2019stabilization}.

The OTD system (\ref{eq:25}) is a set of time-dependent differential equations which must be solved concurrently with the dynamical system (\ref{eq:21}).  The standard approach for infinite-dimensional systems is to discretize (\ref{eq:21}) and (\ref{eq:25}) in space, and advance each using a time-stepping scheme.  The dimension $d$ of the phase space after discretization may be quite large, with the discretized state potentially having thousands of millions of degrees of freedom.  In such cases, computation of the first $r$ OTD modes involves solving $r$ $d$-dimensional differential equations (the OTD equations), plus a $d$-dimensional differential equation for the state itself.  For very large $d$, this procedure is computationally onerous, and alternative approaches might be desirable.

\subsection{Formulation of the learning problem}
\label{sec:22}

For dynamical systems satisfying the assumptions made earlier (autonomy, invertibility, ergodicity, and measure-invariance), the OTD modes asymptotically converge to a set of vectors defined at every point on the attractor \cite{ershov1998concept,goldhirsch1987stability,blanchard2019analytical}.  In other words, in the post-transient regime, $u_i$ only depends on the state $x$, but not on the history of the trajectory or its own initial condition $u_i(t_0)$.  Hence, we may cease to view $u_i$ as being parametrized by $t$, and instead view it as a graph from phase space to tangent space:
\begin{align}
u_i \colon  \mathcal{X} 	&\longrightarrow \mathcal{T}_x \mathcal{X} \nonumber \\
				x 	&\longmapsto    u_i(x).
\label{eq:27}
\end{align}
In this context, the collection $\{u_i(x)\}_{i=1}^r$ has been referred to as the ``stationary Lyapunov basis'' (SLB) at point $x$ (Ref. \onlinecite{ershov1998concept}). 

The existence of the SLB at every point $x$ of the attractor was established separately by Ershov \& Potapov \cite{ershov1998concept} and Goldhirsch et al. \cite{goldhirsch1987stability} as a consequence of the Oseledec theorem \cite{oseledec1968multiplicative}.  The question of uniqueness and continuity with respect to $x$ was also addressed by Ershov \& Potapov \cite{ershov1998concept}.  They showed that for a given $x$, more than one SLB may exist, but only one is stable with respect to perturbations of the underlying state.  So the OTD modes $u_i(x)$ are uniquely determined by the point $x$ in phase space.  They also showed that if the Lyapunov spectrum is not quasi-degenerate (i.e., all Lyapunov exponents are distinct, and there is no ``crossing'' of Lyapunov exponents under small perturbations), then the graph (\ref{eq:27}) is continuous in $x$.  Uniqueness and continuity are important because they allow for the possibility of representing the graph (\ref{eq:27}) as a superposition of smooth basis functions, or as a realization of a Gaussian process.  Once the graph (\ref{eq:27}), or an approximation of it, is known, the Lyapunov exponents can be computed by replacing time-averaging with measure-averaging in (\ref{eq:26}):
\begin{align}
\lambda_i = \int_\mathcal{X} \langle u_i(x), L(x;u_i(x)) \rangle \mathrm{d}\mu(x).
\label{eq:28}
\end{align}

A promising approach is to learn the mapping (\ref{eq:26}) from data.  This requires several ingredients.  First, we assume that we have available a large collection of ``snapshots'' $\{x_n\}_{n=1}^{N}$ for the state.  Each  $x_n$ must belong to the attractor, but not necessarily to the same trajectory, a consequence of the use of measure-averaging.  Second, we assume that we have a mechanism to compute the vector field $F(x_n)$ and the action of the linearized operator $L$ at $x_n$ in any direction $v$.  Third, we need to eliminate the dependence of the OTD system (\ref{eq:25}) on time.  This is done by applying the chain rule to the left-hand side of (\ref{eq:25}), resulting in
\begin{align}
\mathrm{d}u_i(x; F(x)) &= L(x;u_i) - \langle u_i, L(x;u_i) \rangle u_i \nonumber \\
&\quad - \sum_{k=1}^{i-1} [ \langle u_i, L(x;u_k) \rangle + \langle u_k, L(x;u_i) \rangle] u_k,  
\label{eq:29}
\end{align}
where $\mathrm{d}u_i(x; F(x))$ is the G\^ateaux derivative of $u_i$ evaluated at $x$ along the direction $F(x)$.  There is no explicit temporal dependence in (\ref{eq:29}), so that the variable $x$ should no longer be thought of as a point on a particular time-dependent trajectory, but rather as any point on the attractor.  System (\ref{eq:29}) may also be thought of as a partial differential equation on $\mathcal{X}$.  

For computational purposes, it is useful to consider the discretized counterpart of (\ref{eq:29}):
\begin{align}
\nabla_{\!\mathbf{x}} \mathbf{u}_i \, \mathbf{F}(\mathbf{x}) &= \mathbf{L}(\mathbf{x}) \mathbf{u}_i - \langle \mathbf{u}_i, \mathbf{L}(\mathbf{x})  \mathbf{u}_i \rangle \mathbf{u}_i \nonumber \\ 
&\quad - \sum_{k=1}^{i-1} \left[ \langle  \mathbf{u}_i, \mathbf{L}(\mathbf{x}) \mathbf{u}_k \rangle + \langle  \mathbf{u}_k,\mathbf{L}(\mathbf{x}) \mathbf{u}_i \rangle  \right] \mathbf{u}_k, 
\label{eq:210}
\end{align}
where $\mathbf{x}$ and $\mathbf{u}_i$ belong to $\mathbb{R}^d$, $\mathbf{L}(\mathbf{x}) \triangleq \nabla_{\!\mathbf{x}} \mathbf{F}(\mathbf{x})$ is the Jacobian of the vector field $\mathbf{F} \colon \mathbb{R}^d \longrightarrow \mathbb{R}^d$, and $\nabla_{\!\mathbf{x}} \mathbf{u}_i$ is the Jacobian of $\mathbf{u}_i$ with respect to $\mathbf{x}$.  Although not explicitly shown in (\ref{eq:210}), the vector $\mathbf{u}_i$ should be understood as $\mathbf{u}_i(\mathbf{x})$.  We are now in a position to state the learning problem:
\begin{problem}
Given a dataset $\{\mathbf{x}_n\}_{n=1}^{N}$ of snapshots belonging to the attractor $\mathcal{A}$, and a mechanism to compute $\mathbf{F}(\mathbf{x}_n)$ and the action of $\mathbf{L}(\mathbf{x}_n)$, find the collection of graphs $\{\mathbf{x} \longmapsto \mathbf{u}_i(\mathbf{x})\}_{i=1}^r$ that best satisfies (\ref{eq:210}) at every $\mathbf{x}_n$.
\end{problem}
In what follows, we solve the learning problem by a deep-learning approach based on neural networks.

\section{Learning the OTD modes from data}
\label{sec:3}

We will find it convenient to operate in the ``big-data'' regime, so we assume that the dataset used in the learning algorithm contains a very large number of snapshots.  Before we proceed, we reiterate the fundamental assumptions that are made about the data.  The underlying dynamical system from which data is collected should be autonomous, invertible, measure-preserving and ergodic, and should have a non-quasidegenerate Lyapunov spectrum.  As discussed in \S\ref{sec:22}, these assumptions are key to ensure existence, unicity, and continuity of the SLB in phase space.

\subsection{Network architrecture}
\label{sec:31}

To learn the graphs $\{\mathbf{u}_i\}_{i=1}^r$ from the collection of snapshots, we employ a neural-network approach. This is appropriate, because each $\mathbf{u}_i$ is a continuous function of $\mathbf{x}$.  This allows us to leverage the universal approximation theorem \cite{hornik1989multilayer}, which states that any function may be approximated by a sufficiently large and deep neural network.  

Neural networks do suffer from several shortcomings.  They come with heavy computational requirements, are sensitive to the choice of hyper-parameters, and do not always deliver on noisy data.  But the pros largely outweigh the cons.  First, neural networks are known to be more powerful and more expressive than simpler (neighborhood-based or auto-regressive) methods.  Second, as discussed in \S\ref{sec:1}, neural networks have been used successfully to solve partial differential equations, which is essentially what the learning problem proposed in \S\ref{sec:22} amounts to.  Third, neural networks are generally trained by some variant of stochastic gradient descent, and therefore do not suffer from the requirement that the entire dataset be loaded in memory at training time.  With an eye on large-scale deep learning applications, neural networks are therefore the \textit{de facto} correct choice for our learning problem.

In what follows, we refer to the learned OTD modes as the ``deep OTD (dOTD) modes''.

\subsubsection{Overview of the network architrecture}
\label{sec:311}

\begin{figure}[ht!]
\centering
\subfloat[][]{\includegraphics[width=2.75in, clip=true, trim=0 0 0 0]{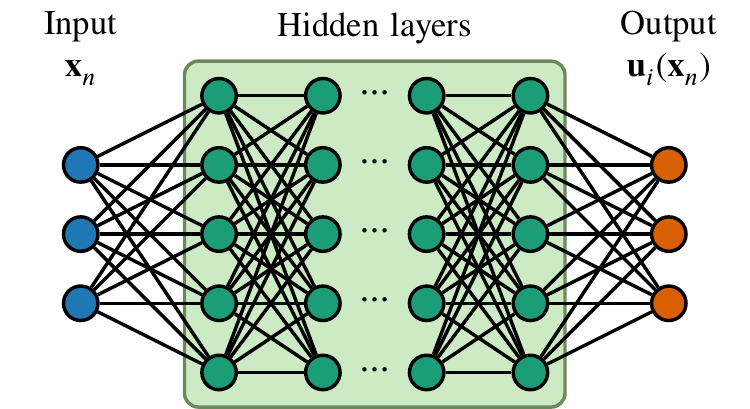}} \qquad
\subfloat[][]{\includegraphics[width=3.4in, clip=true, trim=0 0 0 0]{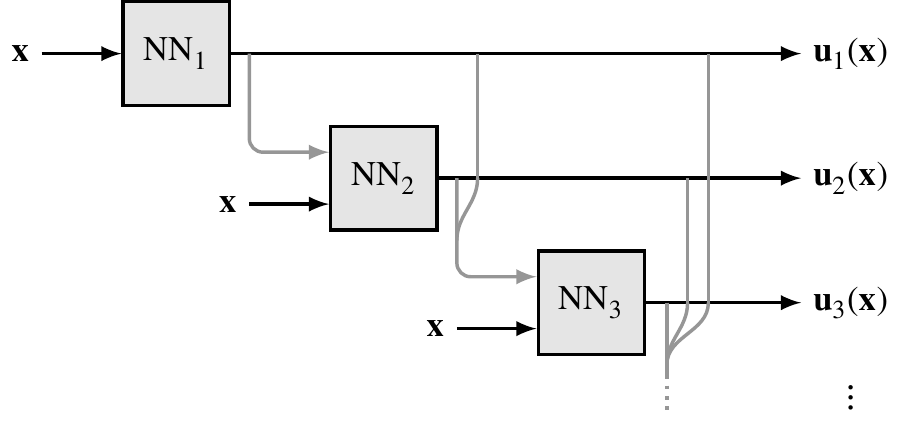}}
\caption{(a) Simplified sketch of the neural-network architecture for the $i$th dOTD mode; and (b) schematic of the sequential approach, where each unit labeled ``$\text{NN}_i$'' is of the type shown in (a), and gray arrows denote that the outputs of the first $i-1$ networks are passed to the $i$th network as dummy inputs.}
\label{fig:0}
\end{figure}

We find it natural to assign to each OTD mode its own neural network $\mathbf{u}_i(\mathbf{x};\bm\theta_i)$, where $\bm\theta_i$ denotes the  parameters (weights and biases) of the $i$th network.  We use the same fully-connected feed-forward architecture with $L$ hidden layers for all OTD modes (figure \ref{fig:0}a).  (The input and output layers are numbered $0$ and $L+1$, respectively.)  The loss function for the $i$th network is specified as 
\begin{widetext}
\begin{align}
\ell_i^\textit{pde}(\bm \theta_i) &=  \frac{1}{N} \sum_{n=1}^N \left\lVert \nabla_{\!\mathbf{x}} \mathbf{u}_i(\mathbf{x}_n;\bm\theta_i) \, \mathbf{F}(\mathbf{x}_n) - \mathbf{L}(\mathbf{x}_n) \mathbf{u}_i(\mathbf{x}_n;\bm\theta_i) + \langle \mathbf{u}_i(\mathbf{x}_n;\bm\theta_i), \mathbf{L}(\mathbf{x}_n)  \mathbf{u}_i(\mathbf{x}_n;\bm\theta_i) \rangle\mathbf{u}_i(\mathbf{x}_n;\bm\theta_i) \phantom{\sum_{k=1}^{i-1}}\right. \nonumber \\
& \qquad\qquad \qquad \left. + \sum_{k=1}^{i-1} \left[ \langle  \mathbf{u}_i(\mathbf{x}_n;\bm\theta_i), \mathbf{L}(\mathbf{x}_n) \mathbf{u}_k(\mathbf{x}_n;\bm\theta_k) \rangle + \langle  \mathbf{u}_k(\mathbf{x}_n;\bm\theta_k),\mathbf{L}(\mathbf{x}_n) \mathbf{u}_i(\mathbf{x}_n;\bm\theta_i) \rangle  \right] \mathbf{u}_k(\mathbf{x}_n;\bm\theta_k)\right\rVert^2,
\label{eq:31}
\end{align}
\end{widetext}
which is nothing more than the residual of the $i$th OTD equation in system (\ref{eq:210}).  We note that any SLB is a global minimizer of $\ell_i^\textit{pde}$, with $\ell_i^\textit{pde}$ being trivially zero.  So this choice of loss function gives rise to no estimation error, and the model is only limited by the hypothesis class (i.e., the set of functions within reach of the neural network for a given number of layers and neurons) and the tolerance specified for the optimization algorithm.  Those are commonly referred to as ``approximation error'' and ``optimization error'', respectively \cite{bottou2018optimization}.  

Equation (\ref{eq:31}) shows that the loss function for the $i$th dOTD mode depends on the first $i-1$ dOTD modes.  This raises the question of the order in which the dOTD modes ought to be learned.  In what follows, we opt for a ``sequential'' approach (figure \ref{fig:0}b), whereby the parameters $\bm\theta_i$ are optimized sequentially (starting with $\bm\theta_1$ and moving down), and the outputs of the first $i-1$ networks are fed into the $i$th network as dummy inputs.  (By ``dummy inputs'', we mean quantities that are fed into the neural network for the sole purpose of computing the loss function.)  We find this architecture to be intuitive because it mimics the triangular structure of the Gram--Schmidt algorithm.  Our numerical experiments suggest that this approach is more stable (compared to other approaches described below), in the sense that it facilitates convergence of the optimization algorithm.  The only issue has to do with error accumulation, arising as a result of using the outcomes of the first $i-1$ optimizations to compute the $i$th dOTD mode.  However, this is easily fixed by tightening the tolerance of the optimization algorithm, or doing multiple passes of training with decreasing tolerance. 

Of course, the ``sequential'' approach is not the only option.  Instead, one could solve the $r$ optimization problems concurrently, whereby the optimization algorithm performs one iteration for each neural network before updating the parameters.  Each iteration would need to be done sequentially (i.e., starting with $\bm\theta_1$, then $\bm\theta_2$, etc), and ``coupling'' between the dOTD modes would be done by passing the first $i-1$ parameter vectors at the current iteration to the $i$th neural network.  Alternatively, one could combine the $r$ loss functions (\ref{eq:31}), and solve for all of the dOTD modes in a single optimization pass using that combined loss function.  However, these two approaches appear to cause difficulty for the optimization algorithm, both in terms of execution speed and its ability to converge.  

Use of the loss function (\ref{eq:31}) alone, which is solely based on the residual of the OTD system, might be problematic for two reasons.  First, as discussed in \S\ref{sec:22}, any SLB is a global minimizer of (\ref{eq:31}), but only one of them is stable (this is the SLB to which the OTD modes converge when computed with a time-marching approach).  So we need a mechanism to ensure that the optimization algorithm converges to the stable SLB, and not to any of the unstable ones.  Second, there may be other (possibly local) minimizers of (\ref{eq:31}) besides SLBs, so we also need a mechanism that prevents the optimization routine from getting trapped in an irrelevant minimum.  

\subsubsection{Ensuring orthonormality of the dOTD modes}
\label{sec:312}

We begin with the question of how to ensure that the optimization algorithm converges to an SLB, with no consideration yet for whether that SLB is stable or not.  We first note that minimization of the loss function (\ref{eq:31}) does not guarantee orthonormality of the resulting dOTD modes.  (For example, the trivial solution $\mathbf{u}_i = \mathbf{0}$ is a global minimizer of (\ref{eq:31}).)  The reason is that the terms responsible for orthonormalizing  the OTD modes in the time-dependent problem (\ref{eq:25}) are no longer sufficient to enforce this constraint.  So orthonormality must be enforced explicitly in the neural network.  This is important because our numerical experiments suggest that the SLBs are the only orthonormal minimizers of (\ref{eq:31}).  In other words, explicitly enforcing orthonormality of the dOTD modes helps the optimization algorithm systematically converge to an SLB, rather than some other irrelevant minimum.

Enforcing orthonormality of the dOTD modes can be realized in two ways.  One approach is to embed Gram--Schmidt orthonormalization immediately after the last layer of the network, so that the dOTD modes are orthonormal by construction.  Another approach is to append to the loss function (\ref{eq:31}) a regularization term,
\begin{align}
\ell_i^{\textit{reg}}(\bm \theta_i)  &=  \frac{1}{N} \sum_{n=1}^N \left\{ \alpha \left(\langle \mathbf{u}_i(\mathbf{x}_n;\bm\theta_i),\mathbf{u}_i(\mathbf{x}_n;\bm\theta_i) \rangle  -1 \right)^2 \right. \nonumber \\
&\qquad \left. +  \sum_{k=1}^{i-1} \beta_k  \langle \mathbf{u}_i(\mathbf{x}_n;\bm\theta_i),\mathbf{u}_k(\mathbf{x}_n;\bm\theta_k) \rangle^2 \right\}.
\label{eq:32}
\end{align}
The first term in the curly brackets enforces normality of $\mathbf{u}_i$, and the second term enforces orthogonality of $\mathbf{u}_i$ and $\mathbf{u}_k$ ($k < i$).  The regularization parameters $\alpha$ and $\beta_k$ should be chosen on a case-by-case basis, which offers less flexibility than the Gram--Schmidt approach.  We note that the regularization approach and the Gram--Schmidt approach have no consequence for the estimation error, because their respective loss functions are exactly zero for any SLB.

In practice, we have found that the Gram--Schmidt approach is more robust than the regularization approach, in the sense that the former requires fewer iterations for the optimization algorithm to converge to an SLB.  We note that each iteration in the Gram--Schmidt approach requires computing the gradients of the Gram--Schmidt layer with respect to the network parameters by back-propagation, which is significantly more expensive than computing the gradients of the regularizing terms in (\ref{eq:32}).  However, this is a burden worth carrying, given that we have found cases in which the regularization approach failed to converge while the Gram--Schmidt approach succeeded, and no cases suggesting otherwise.

\subsubsection{Ensuring uniqueness of the dOTD modes}
\label{sec:313}

Now that we have designed a mechanism ensuring that the optimization algorithm converges to an SLB, we must address the question of how to isolate the stable SLB from all of the unstable ones.  We begin with a simple example that provides insight into this issue.  Consider a case in which the trajectory of interest is a hyperbolic fixed point, denoted by $\mathbf{x}_e$.  Theorem 2.3 in Babaee \& Sapsis \cite{babaee2016minimization} states that any $r$-dimensional eigenspace of the operator $\mathbf{L}(\mathbf{x}_e)$ is an SLB of $\mathbf{x}_e$, and that the only stable SLB is that associated with the $r$ most unstable eigenvalues of $\mathbf{L}(\mathbf{x}_e)$.  In that work, ``stability'' was determined by examining the time-dependent problem governing the evolution of a perturbed SLB.  In the learning problem, however, we have eliminated any notion of temporality from the OTD system.  Hence, for the case of a hyperbolic fixed point, the neural network, in its current manifestation, may converge to any of the $d$-choose-$r$ eigenspaces of $\mathbf{L}(\mathbf{x}_e)$, and not necessarily to the most unstable one.  This is problematic because the OTD modes draw their power from their ability to track directions of greatest instabilities.  Naturally, we wish our learning algorithm to also have this feature.

To make sure that the learning algorithm returns the SLB associated with directions of strongest instabilities, we use a criterion based on Lyapunov exponents.  As discussed in \S\ref{sec:22}, the $i$th Lyapunov exponent $\lambda_i$ can be computed as a measure-average of the Lagrange multiplier $\langle \mathbf{u}_i(\mathbf{x}), \mathbf{L}(\mathbf{x})\mathbf{u}_i(\mathbf{x})\rangle$.  With a finite-size dataset, however, we can do no better than approximating $\lambda_i$ by a finite sum over the data points.  If the dataset is composed of multiple long trajectories initialized on the attractor according to some probability distribution (in general, we want the initial conditions to be independent and identically distributed), we have that
\begin{equation}
\lim_{N \to \infty} \frac{1}{N} \sum_{n=1}^N \langle \mathbf{u}_i(\mathbf{x}_n),  \mathbf{L}(\mathbf{x}_n) \mathbf{u}_i(\mathbf{x}_n) \rangle = \lambda_i,
\label{eq:33}
\end{equation}
where $\mathbf{x}_n$ is the state of the $n$th trajectory after some long time $T$.  The above limiting statement holds only when $T \to \infty$, but in practice we merely require that $T$ be large enough so as to ensure convergence of the distribution of initial conditions to an invariant one.  Equation (\ref{eq:33}) also holds when the dataset is composed of uniformly-sampled snapshots collected in the asymptotic regime of a single long trajectory, in which case (\ref{eq:33}) is equivalent to standard time-averaging.  (Note that in either case the snapshots may be arranged in any arbitrary order.)  Equation (\ref{eq:33}) can be modified to account for the fact that $\mathbf{u}_i$ is modeled as neural network.  We define
\begin{equation}
\hat{\lambda}_i(\bm\theta_i) = \frac{1}{N} \sum_{n=1}^N \langle \mathbf{u}_i(\mathbf{x}_n;\bm\theta_i),  \mathbf{L}(\mathbf{x}_n) \mathbf{u}_i(\mathbf{x}_n;\bm\theta_i) \rangle
\label{eq:34}
\end{equation}
as the ``learned'' Lyapunov exponent associated with the $i$th dOTD mode.  This is the best approximation of $\lambda_i$ available, given the constraints related to finiteness of the dataset and representability of the OTD modes with neural networks.

With (\ref{eq:34}) in hand, we can append to the loss function (\ref{eq:31}) a regularization term, 
\begin{equation}
\ell_i^{\textit{lyap}}(\bm\theta_i) = -\sigma(\hat{\lambda}_i (\bm\theta_i)),
\label{eq:35}
\end{equation}
that penalizes small Lyapunov exponents.  Here, $\sigma$ is a monotonically increasing function on $\mathbb{R}$ that exacerbates differences between the $\hat{\lambda}_i$'s.  Possible choices for $\sigma$ include $\sigma(a) = a$, $a^3$, $\sinh(a)$, and $-\exp(-a)$.  Lyapunov regularization guarantees that the SLB to which the optimization algorithm converges is such that the $\hat{\lambda}_i$'s come in decreasing order, that is, the $i$th dOTD mode picks up the $i$th-largest Lyapunov exponent.  Lyapunov regularization thus ensures that the dOTD modes learned by the neural network coincide with the unique stable solution of the time-dependent OTD system (\ref{eq:25}).

Adding Lyapunov regularization to the loss function (\ref{eq:31}) has the effect of introducing an estimation error, because the value of $\ell_i^{\textit{lyap}}(\bm\theta_i)$ for the optimal $\bm\theta_i$ is generally non-zero, except in very specific cases (for example, if $\sigma(a) = a$ and $\hat{\lambda}_i$ is zero).  No estimation error is a feature worth preserving, because it allows us to focus our attention on the approximation error and the optimization error, thereby facilitating design and optimization of the neural network.  To this effect, we specify an optimization schedule so that Lyapunov regularization is ``switched off'' after a certain number of iterations.  This allows us to ``steer'' the optimization algorithm into a favorable direction initially, while ensuring no estimation error for iterations subsequent to relaxation of Lyapunov regularization.

\subsection{Attractor reconstruction}
\label{sec:32}

The last ingredient needed to make the method fully data-driven is a mechanism to reconstruct the vector field $\mathbf{F}(\mathbf{x}_n)$ and the action of the Jacobian matrix $\mathbf{L}(\mathbf{x}_n)$ from the collection of snapshots $\{\mathbf{x}_n\}_{n=1}^N$.  We note that if the governing equations of the dynamical system are known, then there is no need for such a mechanism because $\mathbf{F}(\mathbf{x}_n)$ and $\mathbf{L}(\mathbf{x}_n)$ can be evaluated directly from the equations of motion.  We also note that if we can only record some observable $f(\mathbf{x})$ of the state, but not the state itself, then we can use delay embedding to reconstruct the attractor, and compute the dOTD modes in the embedded space. (This case will not be considered in this work.)

As discussed in \S\ref{sec:1}, discovery and reconstruction of governing equations from data is an active field of research.  Any of the state-of-the-art methods could be applied to the present problem, each introducing its own level of complexity.   The key issue is that reconstruction of $\mathbf{F}(\mathbf{x})$ can be done offline, regardless of the dimensionality of the system.  In other words, computation of $\mathbf{F}(\mathbf{x}_n)$ for each $\mathbf{x}_n$ may be viewed as a preprocessing step, and therefore does not add to the computational burden related to optimizing the parameters of the neural network.  In what follows, we opt for perhaps the simplest of all approaches.  We assume that the snapshots are sampled along a single long trajectory with a uniform and sufficiently small sampling time-step $\Delta t_s$, so that we may approximate $\mathbf{F}(\mathbf{x}_n)$ as a standard Euler-forward finite difference:
\begin{equation}
\mathbf{F}(\mathbf{x}_n) = \frac{\mathbf{x}_{n+1}-\mathbf{x}_n}{\Delta t_s} + \mathcal{O}(\Delta t_s).
\label{eq:36}
\end{equation}
Higher-order finite-difference formulas (e.g., Adams--Bashforth, Adams--Moulton, or backward differentiation formulas) may be used if higher accuracy is desired.  Finite differences have the advantage of being extremely cheap to compute, even for high-dimensional systems.  Implementation is straightforward, and the requirement that $\Delta t_s$ be small is far from drastic.

To compute the action of the Jacobian matrix $\mathbf{L}(\mathbf{x})$ from data, we employ the classical algorithm proposed independently by Eckmann et al. \cite{eckmann1986liapunov} and Sano \& Sawada \cite{sano1985measurement}.  First, we scan the dataset to identify the $K$ nearest neighbors of each datapoint $\mathbf{x}_n$.  The nearest neighbors of $\mathbf{x}_n$ are defined as those points $\mathbf{x}_{k}$ of the orbit (past or future) that are contained in a ball of radius $\gamma$ centered at $\mathbf{x}_n$:
\begin{equation}
\Vert \mathbf{x}_n - \mathbf{x}_{k} \Vert \leq \gamma, \quad k \in \{1, \dots, K\}.
\label{eq:37}
\end{equation}
If $\gamma$ is sufficiently small, then each vector $\mathbf{v}_{k}^{n} = \mathbf{x}_n - \mathbf{x}_{k}$ may be viewed as a perturbation vector from the reference orbit.  We therefore have
\begin{equation}
\frac{\mathbf{v}_{k+1}^{n+1} - \mathbf{v}_{k}^{n}}{\Delta t_s} =  \mathbf{L}(\mathbf{x}_n)  \mathbf{v}_{k}^{n} + \mathcal{O}(\Delta t_s),
\label{eq:38}
\end{equation}
which allows us to compute the action of the Jacobian matrix $\mathbf{L}(\mathbf{x}_n)$ on the vectors $\{\mathbf{v}_{k}^{n}\}_{k=1}^K$.   Now, the critical step is to note that the vectors $\{\mathbf{v}_{k}^{n}\}_{k=1}^K$ belong to the tangent space at point $\mathbf{x}_n$, and so do the OTD modes.  (In fact, the OTD modes form a basis of that space when $r=d$.)  So if we stack up the vectors $\{\mathbf{v}_{k}^{n}\}_{k=1}^K$ into a matrix $\mathbf{V}_n \in \mathbb{R}^{d \times K}$, then the least-square fit
\begin{equation}
\mathbf{u}_i(\mathbf{x}_n; \bm \theta_i) \approx \mathbf{V}_n \mathbf{V}_n ^\dagger \mathbf{u}_i(\mathbf{x}_n; \bm \theta_i)
\label{eq:39}
\end{equation}
should be a reasonably good approximation for the dOTD modes.  Here, $\mathbf{V}_n^\dagger$ is the pseudo-inverse of $\mathbf{V}_n$.  (We note that the least-square approach allows for $K$ exceeding the dimension of the phase space $d$.)  From (\ref{eq:39}), we can compute the action of the linearized operator on the dOTD modes as
\begin{equation}
\mathbf{L}(\mathbf{x}_n) \mathbf{u}_i(\mathbf{x}_n; \bm \theta_i) \approx \Delta\!\mathbf{V}_n \mathbf{V}_n^\dagger \mathbf{u}_i(\mathbf{x}_n; \bm \theta_i),
\label{eq:310}
\end{equation}
where $\Delta\!\mathbf{V}_n$ is a $d$-by-$K$ matrix with columns $(\mathbf{v}_{k+1}^{n+1} - \mathbf{v}_{k}^{n})/\Delta t_s$.  Equation (\ref{eq:310}) requires no information other than the snapshot data, and can be used to evaluate the loss function (\ref{eq:31}).

For low-dimensional systems, it is possible to form the matrix $\Delta\!\mathbf{V}_n \mathbf{V}_n^\dagger \in \mathbb{R}^{d \times d}$ explicitly, and store it in memory.  This computation can be done offline for every datapoint $\mathbf{x}_n$, and consequently the computational burden associated with reconstruction is nil for the neural network.  For high-dimensional systems, however, forming and storing $\Delta\!\mathbf{V}_n \mathbf{V}_n^\dagger$ is not possible, so (\ref{eq:310}) must be computed online (i.e., every time the loss function (\ref{eq:31}) is evaluated), which introduces an additional cost.  Another aggravating issue is that the complexity of most nearest-neighbor-search algorithms grows exponentially with the dimension of the state, making the above approach intractable for $d$ greater than about 25 (Ref. \onlinecite{dobkin1976multidimensional}).  The curse of dimensionality thus requires that we pursue a different strategy.

\subsection{Learning in a high-dimensional phase space}
\label{sec:33}

For high-dimensional systems, the neural-network approach is intractable for two reasons. First, there is the issue of reconstructing the Jacobian matrix (or, equivalently, its action on the dOTD modes), which was discussed in \S\ref{sec:32}.  Second, to evaluate the term $\nabla_{\!\mathbf{x}} \mathbf{u}_i(\mathbf{x}_n;\bm\theta_i)$ appearing in (\ref{eq:31}), we must compute the gradient of the dOTD modes with respect to the state vector, resulting in a $d$-by-$d$ matrix.  For large $d$, this computation is virtually hopeless.  So to make the neural-network approach applicable to high-dimensional systems, we proceed to an order-reduction of the phase space.  

If $\mathbf{x}$ arises from discretizing a partial differential equation (PDE) defined in a domain $\Omega$, then one approach is to randomly select $M$ points in that domain according to some probability distribution.  Each snapshot $\mathbf{x}_n$ then has dimension $M$, with $M$ presumably much smaller than $d$.  When $\mathbf{x}$ contains nodal values of the state, this approach amounts to randomly excising $M$ entries from each $\mathbf{x}_n$.  (The excised entries need not be the same for all the snapshots.)  Random sampling has been used successfully in a number of problems related to deep learning of PDEs \cite{sirignano2018dgm,raissi2018deep}, largely because it has the advantage of being a ``mesh-free'' approach.  However, applicability of this method to the present problem is limited, because the algorithm for attractor reconstruction requires that the sampled points be the same for all snapshots.  As a result, we may need a large number of sampled points (with $M$ possibly on the order of $d$) to faithfully capture the dynamics.  If $M$ is reasonably small, however, the dOTD modes can be learned at the sampled points, and subsequently reconstructed over the entire domain using any standard interpolation algorithm.

To avoid these difficulties, we use an approach based on Galerkin projection.  We assume that any state on the attractor can be represented as a superposition of proper-orthogonal-decomposition (POD) modes,
\begin{equation}
\mathbf{x} - \bar{\mathbf{x}} = \bm\Phi \bm\xi,
\label{eq:311}
\end{equation}
where $\bar{\mathbf{x}}$ is the mean flow, $\bm\Phi \in \mathbb{R}^{d \times n_s}$ contains the first $n_s$ POD modes, and $\bm\xi \in \mathbb{R}^{n_s}$ contains the corresponding POD coefficients.  Measure-invariance and ergodicity of the attractor allow use to view $\bm\xi$ as a function of time or as a function of the state, so that we may use $\bm\xi(t)$ and $\bm\xi(\mathbf{x})$ interchangeably.  Each POD mode can be computed directly from data by the method of snapshots \cite{sirovich1987turbulence}, and the $i$th POD coefficient can be obtained by projecting $\mathbf{x} - \bar{\mathbf{x}}$ onto the $i$th POD mode.  The number of retained POD modes is determined by examining the cumulative energy of the POD eigenvalues.  In general, $n_s$ is selected so as to account for at least 95\% of the total energy.  We also note that implicit in (\ref{eq:311}) is a one-to-one correspondence between snapshots in $\mathbf{x}$-space ($\mathbf{x}_n$) and snapshots in $\bm\xi$-space ($\bm\xi_n$).  This allows us to view any function of $\mathbf{x}_n$ as a function of $\bm\xi_n$, and vice versa.

In the POD subspace, the dynamics obeys
\begin{equation}
\dot{\bm\xi} =   \bm\Phi^\mathsf{T} \mathbf{F}(\bar{\mathbf{x}} + \bm\Phi \bm\xi) \triangleq \mathbf{G}(\bm\xi),
\label{eq:312}
\end{equation}
so in principle we could use the neural-network approach to learn the OTD modes in the $\bm\xi$-subspace.  We would simply apply the method proposed in \S\ref{sec:31} and \S\ref{sec:32} with $\bm\xi$ in place of $\mathbf{x}$, $\mathbf{G}$ in place of $\mathbf{F}$, and $\nabla_{\!\bm\xi}\mathbf{G}$ in place of $\mathbf{L}$; then project the dOTD modes learned in the $\bm\xi$-subspace back to the full space, resulting in $\bm\Phi \mathbf{u}_i(\bm\xi)$.  However, this approach is flawed, because by construction it assumes that the OTD modes live in the same subspace as the state itself (that is, the POD subspace spanned by the columns of $\bm\Phi$).  In reality, the OTD modes belong the tangent space at point $\mathbf{x}$, whose principal directions have no reason to coincide with that of the state on the attractor.  This inconsistency persists regardless of the number of POD modes used in (\ref{eq:311}), so that resorting to very large $n_s$ does not solve the problem.  

All is not lost, though, since we can use different projection subspaces for the state and the OTD modes:
\begin{subequations}
\begin{gather}
\mathbf{x} - \bar{\mathbf{x}} = \bm\Phi \bm\xi(\mathbf{x}), \\
\mathbf{u}_i(\mathbf{x}) = \bm\Psi(\mathbf{x}) \bm \mu_i(\mathbf{x}),
\end{gather}%
\label{eq:313}%
\end{subequations} 
where $\bm\Psi(\mathbf{x})\in \mathbb{R}^{d \times n_t}$ is a reduced orthonormal basis of the tangent space at $\mathbf{x}$, and $\bm \mu_i(\mathbf{x}) \in \mathbb{R}^{n_t}$ contains the basis coefficients.    We note that the number of POD modes ($n_s$) and columns of $\bm\Psi$ ($n_t$) need not be the same, which allows for the possibility of learning the OTD modes in a subspace bigger than the POD subspace.  If $\bm\Psi(\mathbf{x}_n)$ is a good approximation of the tangent space at $\mathbf{x}_n$, then the $K$ nearest neighbors of $\mathbf{x}_n$ satisfy
\begin{equation}
\mathbf{x}_{k} \approx \mathbf{x}_n + \bm\Psi(\mathbf{x}_n) \mathbf{a}_{nk}, \quad k \in \{1, \dots, K\},
\label{eq:314}
\end{equation}
where $\mathbf{a}_{nk}$ is a matrix of coefficients.  To compute $\bm\Psi(\mathbf{x}_n)$ from data, we solve the minimization problem
\begin{equation}
\bm\Psi(\mathbf{x}_n) =\operatorname*{argmin}_{\bm\Psi,\,\mathbf{a}} \Vert \mathbf{x}_{k} -( \mathbf{x}_n + \bm\Psi \mathbf{a}) \Vert^2,
\label{eq:315}
\end{equation}
which is tantamount to a principal component analysis (PCA) of the set of nearest neighbors $\{\mathbf{x}_{k}\}_{k=1}^K$.  In other words, the columns of $\bm\Psi(\mathbf{x}_n)$ are the leading $n_t$ POD modes of the $K$ nearest neighbors of $\mathbf{x}_n$; we refer to them as the ``tangent POD (tPOD) modes''.   We note that this approach has been used as the basis for a number of manifold learning algorithms that involve reconstruction of tangent space from data \cite{roweis2000nonlinear,zhang2004principal}.

Now that we have available two low-dimensional representations for the state and the OTD modes, the learning problem reduces to finding the collection of graphs
\begin{align}
\bm \mu_i \colon  \mathbb{R}^{n_s} 	&\longrightarrow \mathbb{R}^{n_t} \nonumber \\
				\bm \xi 		&\longmapsto    \bm \mu_i(\bm \xi),
\label{eq:316}
\end{align}
given a dataset of reduced states $\{\bm \xi_n \}_{n=1}^N$ and tangent bases $\{\bm\Psi(\bm \xi_n) \}_{n=1}^N$.  To derive an equation for $\bm \mu_i$, we substitute (\ref {eq:313}a,b) into (\ref{eq:210}) and arrive at
\begin{align}
\nabla_{\!\bm\xi} \bm\mu_i \, \mathbf{G}(\bm\xi) &= \mathbf{L}_{\bm\Psi} \bm\mu_i - \langle \bm\mu_i, \mathbf{L}_{\bm\Psi}  \bm\mu_i \rangle \bm\mu_i \nonumber \\
&\quad - \sum_{k=1}^{i-1} \left[ \langle  \bm\mu_i, \mathbf{L}_{\bm\Psi} \bm\mu_k \rangle + \langle  \bm\mu_k,\mathbf{L}_{\bm\Psi} \bm\mu_i \rangle  \right] \bm\mu_k \nonumber \\
&\quad - \bm\Psi^\intercal \dot{\bm\Psi} \bm\mu_i.  
\label{eq:317}
\end{align}
Three remarks are in order.  First, we note that (\ref{eq:317}) is a system of $n_t$-dimensional differential equations, much less expensive to solve that the original system of $d$-dimensional differential equations.  Second, we note the presence of an additional term on the right-hand side of (\ref{eq:317}) arising from the dependence of the tPOD modes on the state $\mathbf{x}$.  Third, we have defined the reduced operator 
\begin{equation}
\mathbf{L}_{\bm\Psi} = \bm\Psi(\mathbf{x})^\intercal \mathbf{L}(\mathbf{x}) \bm\Psi(\mathbf{x}) \in \mathbb{R}^{n_t \times n_t},
\label{eq:318}
\end{equation}
which is the projection of the high-dimensional operator $\mathbf{L}(\mathbf{x})$ onto the reduced basis $\bm\Psi(\mathbf{x})$.  Equation (\ref{eq:318}) provides additional insight as to why the ``naive'' approach described two paragraphs earlier was not a good one.  That approach was equivalent to having $\bm\Psi(\mathbf{x}) = \bm\Phi$ for every point $\mathbf{x}$ on the attractor, leading to $\mathbf{L}_{\bm\Psi} = \bm\Phi^\mathsf{T} \mathbf{L}(\mathbf{x})\bm\Phi$.  But this is a poor approximation for $\mathbf{L}(\mathbf{x})$, because $\bm\Phi$ is a reduced basis of the phase space, not the tangent space. 

To compute the reduced operator $\mathbf{L}_{\bm\Psi}(\mathbf{x}_n)$ from data, we use an approach similar to that proposed in \S\ref{sec:32}.  We first recall that (\ref{eq:38}) provides a mechanism to compute the action of $\mathbf{L}(\mathbf{x}_n)$ on a collection of perturbation vectors $\{\mathbf{v}_k^n\}_{k=1}^K$ at point $\mathbf{x}_n$.  But the columns of $\bm\Psi(\mathbf{x}_n)$ are linear combinations of these perturbation vectors, as per the POD construction.  Therefore, we may write $\bm\Psi(\mathbf{x}_n) =  \mathbf{V}_n \bm\kappa_n$, where $\bm\kappa_n \in \mathbb{R}^{K \times n_t}$ is a matrix of coefficients.  This leads to 
\begin{equation}
\mathbf{L}_{\bm\Psi}(\mathbf{x}_n) \approx \bm\kappa_n^\intercal  \mathbf{V}_n^\intercal \Delta\!\mathbf{V}_n \bm\kappa_n.
\label{eq:319}
\end{equation}
The operator $\mathbf{L}_{\bm\Psi}(\mathbf{x}_n)$ has the advantage of being low-dimensional, so it can be computed offline and stored in memory, along with the POD-reduced vector field $\mathbf{G}(\bm\xi_n)$.  

The last term on the right-hand side of (\ref{eq:317}), however, is problematic because it involves the temporal derivative of the reduced tangent basis $\bm\Psi$.  We could use the chain rule and write $\dot{\bm\Psi} = \nabla_{\!\bm\xi} \bm\Psi \, \mathbf{G}(\bm\xi)$, but the gradient $\nabla_{\!\bm\xi} \bm\Psi$ is expensive to compute.  Another option is to use a finite-difference formula in the spirit of (\ref{eq:36}):
\begin{equation}
\dot{\bm\Psi}(\mathbf{x}_{n}) = \frac{\bm\Psi(\mathbf{x}_{n+1})- \bm\Psi(\mathbf{x}_{n})}{\Delta t_s} + \mathcal{O}(\Delta t_s).
\label{eq:320}
\end{equation}
If $\Delta t_s$ is sufficiently small, then the tPOD modes smoothly deform from $\mathbf{x}_{n}$ to $\mathbf{x}_{n+1}$, so (\ref{eq:320}) is a good approximation.  (We note in passing that continuity of $\mathbf{u}_i(\mathbf{x})$ requires continuity of both $\bm\Psi(\mathbf{x})$ and $\bm\mu_i(\mathbf{x})$.)  With this approach, the term $\bm\Psi^\intercal \dot{\bm\Psi}$ can be computed offline, and passed to the neural network as a dummy input.

Finally, we address the question of whether the ``local'' tangent bases $\bm\Psi(\mathbf{x})$ could be combined into a larger ``global'' subspace that does not depend on $\mathbf{x}$.  Mathematically, this is equivalent to seeking a reduced basis of the tangent bundle
\begin{equation}
\mathcal{T}\mathcal{X} = \bigsqcup_{x \in \mathcal{X}} \mathcal{T}_x \mathcal{X},
\end{equation}
where $\bigsqcup$ is the disjoint union operator.  Such a construct would have the benefit of eliminating the last term on the right-hand side of (\ref{eq:317}).  Also, it would be aesthetically more appealing, because the OTD modes would be learned in a common subspace, regardless of the point at which they are computed.  To construct a reduced basis of $\mathcal{T}\mathcal{X}$, one can perform POD on the set of tangent bases $\{\bm\Psi(\mathbf{x}_n) \}_{n=1}^N$.  This results in a set of ``bundle modes'', denoted by $\bm\Pi \in \mathbb{R}^{d \times n_b}$, whose span is the best $n_b$-dimensional approximation of the tangent bundle $\mathcal{T}\mathcal{X}$.  The expectation is that the number of bundle modes $n_b$, although generally greater than the number of local tPOD modes $n_t$, will still be much smaller than $d$.  The dOTD modes are sought as vectors in the bundle space, that is, $\mathbf{u}_i(\mathbf{x}) = \bm \Pi \bm \rho_i(\mathbf{x})$, where the coefficients $\bm \rho_i \in \mathbb{R}^{n_b}$ satisfy
\begin{align}
\nabla_{\!\bm\xi} \bm\rho_i \, \mathbf{G}(\bm\xi) &= \mathbf{L}_{\bm\Pi} \bm\rho_i - \langle \bm\rho_i, \mathbf{L}_{\bm\Pi}  \bm\rho_i \rangle \bm\rho_i \nonumber \\
&\quad - \sum_{k=1}^{i-1} \left[ \langle  \bm\rho_i, \mathbf{L}_{\bm\Pi} \bm\rho_k \rangle + \langle  \bm\rho_k,\mathbf{L}_{\bm\Pi} \bm\rho_i \rangle  \right] \bm\rho_k,  
\label{eq:322}
\end{align}
and $\mathbf{L}_{\bm\Pi}$ is defined as 
\begin{equation}
\mathbf{L}_{\bm\Pi} = \bm\Pi^\intercal \mathbf{L}(\mathbf{x}) \bm\Pi \in \mathbb{R}^{n_b \times n_b}.
\label{eq:323}
\end{equation}
For $n_b$ not too large, the reduced operator $\mathbf{L}_{\bm\Pi}$ may be computed offline and stored in memory. 

Applicability of the tangent-bundle method is limited to cases in which the governing equations of the dynamical system are available.  This is because $\mathbf{L}_{\bm\Pi}$ cannot be computed solely from snapshot data, except when the linearized operator is known $\mathbf{L}(\mathbf{x})$ explicitly.  To see this, we first recognize that by construction, each bundle mode is a linear combination of the columns of $\bm \Upsilon = \begin{bmatrix} \mathbf{V}_1, \dots, \mathbf{V}_n \end{bmatrix}$.  So evaluation of (\ref{eq:323}) from data is conditioned on our ability to compute $\mathbf{L}(\mathbf{x}_n) \bm\Upsilon$ or, equivalently, each member of the set $\{\mathbf{L}(\mathbf{x}_n) \mathbf{V}_m \}_{m=1}^n$.  But there is no mechanism to compute $\mathbf{L}(\mathbf{x}_n) \mathbf{V}_m$ from data when $m \neq n$.  (Equation (\ref{eq:38}) holds only for $m=n$.)  Thus, evaluation of (\ref{eq:323}) requires explicit knowledge of the linearized operator and, in turn, the governing equations.  Because of this restriction, this approach will not be pursued any further in this work.

\subsection{Implementation}

We conclude this section with a few words on implementation.  The neural network is built from scratch in Python.  We use Autograd \cite{maclaurin2015autograd} for automatic differentiation.  The activation function between hidden layers is the hyperbolic tangent, although other choices (e.g., sigmoid function or swish function) are possible \cite{ramachandran2017searching}.  The activation function for the output layer is a linear function.  As discussed in \S\ref{sec:312}, the last layer of the neural network is followed by a Gram--Schmidt layer.  The weights are initialized according to Xavier initialization \cite{glorot2010understanding}.  For Lyapunov regularization, we use $\sigma = \sinh$.

We use Adam \cite{kingma2014adam} to solve the optimization problem.  In its current manifestation, the code supports mini-batching, although caution must be exercised when selecting the batch size in order not to deteriorate the accuracy of (\ref{eq:34}).  (The results presented in \S\ref{sec:4} do not use mini-batching.)  We have found no need for specifying learning rate schedules in the optimizer.  Different stopping criteria may be used, depending on how the vector field and linearized operator are computed.  If $\mathbf{F}(\mathbf{x}_n)$ and $\mathbf{L}(\mathbf{x}_n)$ are evaluated from the governing equations, and the hypothesis class is reasonably large, then optimization may be terminated when the loss function (\ref{eq:31}) is smaller than a user-specified tolerance.  (As a result, the overall error is dominated by the approximation error.)  If $\mathbf{F}(\mathbf{x}_n)$ and $\mathbf{L}(\mathbf{x}_n)$ are reconstructed from data, then it is preferable to terminate optimization after a user-specified number of iterations, in recognition of the fact that the reconstruction process is approximate, and therefore introduces an estimation error.

To generate the dataset, we consider a single long trajectory of the dynamical system, rather than multiple shorter trajectories with distinct initial conditions.  The dataset is constructed by collecting equally-spaced snapshots (with sampling time $\Delta t_s$) along that long trajectory.  (For the results presented in \S \ref{sec:4}, $\Delta t_s$ coincides with the time-step size $\Delta t$ used to advance the dynamical system.)  For each snapshot, we reconstruct the vector field and linearized operator using the \texttt{KNeighborsClassifier} implemented in \texttt{scikit-learn}.  We note that the nearest-neighbor search is carried out over the entire collection of snapshots, including those collected in the transient regime, so as to improve accuracy of the reconstruction algorithm \cite{eckmann1986liapunov}.  For high-dimensional systems, the nearest-neighbor search is conducted in the POD subspace to alleviate computational cost.  (By that, we mean that we merely extract time stamps for the nearest neighbors in $\bm\xi$-space, and then use the corresponding snapshots in $\mathbf{x}$-space in the reconstruction algorithm.)  In the examples presented in \S\ref{sec:4}, the vector field and linearized operator are formed explicitly, stored in memory, and passed to the neural network as dummy inputs.  This is possible either because the dynamical system is low-dimensional, or because we first proceeded to a reduction of the dynamics using the Galerkin approach described in \S\ref{sec:33}.

Upon reconstruction of $\mathbf{F}(\mathbf{x}_n)$ and $\mathbf{L}(\mathbf{x}_n)$, we discard from the dataset those snapshots that were collected in the transient regime of the long trajectory.  As discussed in \S \ref{sec:311}, this step is critical to ensure validity of the measure-averaging operation.  The resulting ``truncated'' dataset is then split into training, validation, and test sets.  The training set is used to optimize the weights of the neural network, and the test set to evaluate its predictive capability.  The validation set is used to tune the hyper-parameters of the neural network, as described in Ref. \onlinecite{goodfellow2016deep}.  (For a given dynamical system, we consider a range of random combinations of hyper-parameters, and select the architecture for which the validation error is the smallest.)  The validation set is also used to tune the learning rate for the Adam optimizer, the schedule for Lyapunov regularization, and the number of nearest neighbors used in the reconstruction step.

The code is available on GitHub \cite{blanchard2019github}.

\section{Results}
\label{sec:4}

\subsection{Performance metrics}
\label{sec:40a}

To evaluate the accuracy of the learning algorithm, we consider two performance metrics.  The first metric is the PDE loss $\ell_i^{\textit{pde}}$ defined in (\ref{eq:31}), which quantifies the extent to which the neural network satisfies the OTD equations (\ref{eq:210}).  We use this metric as our primary guide in our search for appropriate hyper-parameters, because a small PDE loss on training and validation data is a key prerequisite for generalizability on test data.  

However, as discussed in \S \ref{sec:311}, the PDE loss does not discriminate between stable and unstable SLBs. So we supplement it with
\begin{equation}
d_i(\mathbf{x}) = 1- \left| \langle \mathbf{u}_i^\textit{deep}(\mathbf{x}), \mathbf{u}_i^\textit{num}(\mathbf{x}) \rangle \right|,
\label{eq:41}
\end{equation}
where $\mathbf{u}_i^\textit{deep}$ denotes the $i$th dOTD mode, and $\mathbf{u}_i^\textit{num}$ denotes the $i$th OTD mode computed by direct numerical integration of (\ref{eq:25}).  The distance $d_i$ takes values between 0 and 1, with the former indicating that $\mathbf{u}_i^\textit{deep}$ and $\mathbf{u}_i^\textit{num}$ coincide, and the latter indicating that $\mathbf{u}_i^\textit{deep}$ and $\mathbf{u}_i^\textit{num}$ are orthogonal.  We could also use the distance between the subspaces $\{\mathbf{u}_i^\textit{deep}\}_{i=1}^r$ and $\{\mathbf{u}_i^\textit{num}\}_{i=1}^r$, but that measure, unlike (\ref{eq:41}), assigns the same score to all the SLBs, regardless of stability (when $r=d$).

If the OTD modes are learned in a reduced subspace, then it is useful to compute 
\begin{equation}
d_i^{\bm\Theta}(\mathbf{x}) = 1- \left| \langle \mathbf{u}_i^\textit{deep}(\mathbf{x}),  \bm\Theta \bm\Theta^\mathsf{T} \mathbf{u}_i^\textit{num}(\mathbf{x})\rangle \right|,
\label{eq:42}
\end{equation}
where $\bm\Theta$ is a placeholder for $\bm \Psi(\mathbf{x})$ or $\bm \Pi$, depending on whether the ``local'' or ``bundle'' approach is used.  The above is equivalent to computing the distance function in the reduced subspace in which the modes are learned.  This quantity is important, because dimensionality reduction of the tangent space (or tangent bundle) introduces an additional error over which the neural network has no control.  Thus, situations may arise in which $d_i^{\bm\Theta}$ is small, but $d_i$ is large.  Should that occur, a quick fix is to increase the dimension of the reduced subspace $\bm\Theta$.

For each example considered below, we report training, validation, and testing error as measured by the two metrics $\ell_i^{\textit{pde}}$ and $d_i$.  (For the latter, we report its measure-average, $\bar{d}_i$, over the dataset considered.)  These numbers have been averaged over ten learning experiments.  Each learning experiment corresponds to a different (random) initialization of the weights $\bm\theta_i$.  The hyper-parameters, training set, validation set, and test set are kept the same from one learning experiment to the next.  These results were found to be robust to small changes in hyper-parameters and dataset selection.  

\subsection{Examples}
\label{sec:40b}

\subsubsection{Low-dimensional nonlinear system}
\label{sec:41}

We begin with a three-dimensional nonlinear system that was proposed by Noack et al. \cite{noack2003hierarchy} as a testbed for investigating Hopf bifurcations in laminar bluff-body flows.  We choose this system because it provides a good illustration for many of the comments made in \S\ref{sec:2} and \S\ref{sec:3}.  The governing equations are given by
\begin{subequations}
\begin{gather}
\dot{x} = \mu x - y - x z, \\
\dot{y} = \mu y + x - y z, \\
\dot{z} = - z + x^2 + y^2,
\end{gather}
\label{eq:43}%
\end{subequations}%
with $\mu$ a positive constant.  The system admits a linearly unstable fixed point ($x=y=z=0$), and a stable periodic solution,
\begin{equation}
x=\sqrt{\mu} \cos t, \quad y=\sqrt{\mu} \sin t, \quad z = \mu,
\label{eq:44}
\end{equation}
which defines a limit cycle of radius $\sqrt{\mu}$ in the $z=\mu$ plane.  As noted by Noack et al. \cite{noack2003hierarchy}, the limit cycle is asymptotically and globally stable.

For $\mu \leq 1/8$, it is possible to derive analytical expressions for the OTD modes on the limit cycle:
\begin{subequations}
\begin{equation}
\mathbf{u}_1 = \begin{bmatrix}  -\sin t \\ \cos t \\ 0 \end{bmatrix}, ~~ \mathbf{u}_2 = \begin{bmatrix}  -a \cos t \\ -a \sin t \\ -b \end{bmatrix}, ~~ \mathbf{u}_3 = \begin{bmatrix}  -b \cos t \\ -b \sin t \\ a \end{bmatrix},
\label{eq:45a}
\end{equation}
where 
\begin{equation}
a = \frac{1+\sqrt{1-8\mu}}{4 \sqrt{\mu}} b, \quad
b = \sqrt{ \frac{1+4\mu-\sqrt{1-8\mu}}{2(1+\mu)}}.
\label{eq:45b}
\end{equation}
\end{subequations}
These are the modes to which solutions of the OTD equations converge when computed with a time-stepping approach.  (To the best of our knowledge, this is the first time that exact (non-asymptotic) expressions have been reported for unsteady OTD dynamics.)  Each $\mathbf{u}_i$ may be expressed as a function of the state $\mathbf{x} = \begin{bmatrix} x & y & z \end{bmatrix}^\mathsf{T}$ as follows:
\begin{gather}
\mathbf{u}_1(\mathbf{x}) = \frac{\mathbf{F}(\mathbf{x})}{\Vert\mathbf{F}(\mathbf{x})\Vert}, \quad \mathbf{u}_2(\mathbf{x})  = -\frac{a}{\sqrt{\mu}} \mathbf{x} - \begin{bmatrix}  0 \\ 0 \\ b + a/\sqrt{\mu} \end{bmatrix}, \nonumber \\
\mathbf{u}_3(\mathbf{x})  = -\frac{b}{\sqrt{\mu}} \mathbf{x} + \begin{bmatrix}  0 \\ 0 \\ a- b/\sqrt{\mu} \end{bmatrix}.
\label{eq:46}
\end{gather}
The ordered set $\{\mathbf{u}_1(\mathbf{x}),\mathbf{u}_2(\mathbf{x}),\mathbf{u}_3(\mathbf{x})\}$ is the unique stable SLB at point $\mathbf{x}$ on the limit cycle, although other unstable SLBs exist.  If we define 
\begin{equation}
a^\pm = \frac{1\pm\sqrt{1-8\mu}}{4 \sqrt{\mu}} b^\pm, \quad
b^\pm = \sqrt{ \frac{1+4\mu\pm\sqrt{1-8\mu}}{2(1+\mu)}},
\label{eq:47}
\end{equation}
then any of the ordered triples $\{\mathbf{u}_1, \mathbf{u}_2^\pm, \mathbf{u}_3^\pm\}$, $\{\mathbf{u}_2^\pm, \mathbf{u}_1, \mathbf{u}_3^\pm\}$, and $\{\mathbf{u}_2^\pm, \mathbf{u}_3^\pm, \mathbf{u}_1\}$ are solutions to the OTD equations, but only $\{\mathbf{u}_1, \mathbf{u}_2^+, \mathbf{u}_3^+\}$ is stable.  

We consider the case $\mu=0.1$, and use the neural-network approach to learn the graphs $\{ \mathbf{x} \longmapsto \mathbf{u}_i(\mathbf{x})\}_{i=1}^3$ from data.  We begin by generating a long trajectory initiated close to the limit cycle.  This trajectory is computed by a third-order Adams--Bashforth method with time-step size $\Delta t =0.01$ for a total duration of $T=50$ time units.  We record snapshots at every time step, resulting in an initial dataset comprised of 5000 points.  For each datapoint, we reconstruct the vector field and linearized operator using its seven nearest neighbors.  We then discard the first 1000 datapoints corresponding to the transient regime.  The resulting dataset is comprised of 4000 snapshots recorded in the interval $10 \leq t \leq 50$, spanning about six periods in the asymptotic regime of the long trajectory.  We then divide up that dataset into three.  The training set is comprised of 10 equally-spaced points over the period $20 \leq t \leq 20+2\pi$; the validation test of 629 equally-spaced points over the period $10 \leq t \leq 10+2\pi$; and the training set of the remaining points.  (We use the same dataset splitting to train the three neural networks.)  The neural network is composed of one hidden layer with 40 neurons, and the learning rate for the Adam optimizer is set to 0.04.  Lyapunov regularization is active for the first 1000 iterations, and turned off for the remainder of the optimization.  The dOTD modes are learned sequentially, with the maximum number of iterations specified as 3000.  For these parameters, the training, validation, and test errors are shown in table \ref{tab:1}. 

\begin{table}[ht!]
\caption{Empirical validation results for the low-dimensional nonlinear system with $\mu=0.1$ and hyper-parameters given in the text.
\label{tab:1}} 
\begin{ruledtabular} 
\begin{tabular}{c|cc|cc|cc}
 	& \multicolumn{2}{c|}{Training set} 	& \multicolumn{2}{c|}{Validation set}			& \multicolumn{2}{c}{Test set}	\\ 
$i$	& $\ell_i^{\textit{pde}}$	& $\bar{d}_i$	& $\ell_i^{\textit{pde}}$	& $\bar{d}_i$
	& $\ell_i^{\textit{pde}}$	& $\bar{d}_i$ \\
1	& $1.2 \cdot 10^{-6}$		& $1.1 \cdot 10^{-6}$		& $2.7 \cdot 10^{-6}$ &	$9.6 \cdot 10^{-7}$		& $2.5 \cdot 10^{-6}$	 	&	$9.6 \cdot 10^{-7}$ 	\\
2	& $7.9 \cdot 10^{-6}$		& $1.2 \cdot 10^{-4}$		& $3.9 \cdot 10^{-4}$	 &	$1.3 \cdot 10^{-4}$		& $1.9 \cdot 10^{-4}$		& 	$1.4 \cdot 10^{-4}$	\\
3	& $8.2 \cdot 10^{-6}$		& $1.2 \cdot 10^{-4}$		& $3.9 \cdot 10^{-4}$	 &	$1.3 \cdot 10^{-4}$		& $1.9 \cdot 10^{-4}$	 	& 	$1.4 \cdot 10^{-4}$	
\end{tabular} 
\end{ruledtabular} 
\end{table}

\begin{figure}[ht!]
\centering
\subfloat[][]{\includegraphics[width=1.65in]{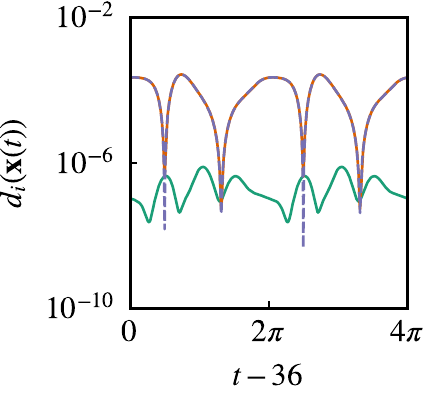}} ~
\subfloat[][]{\includegraphics[width=1.65in]{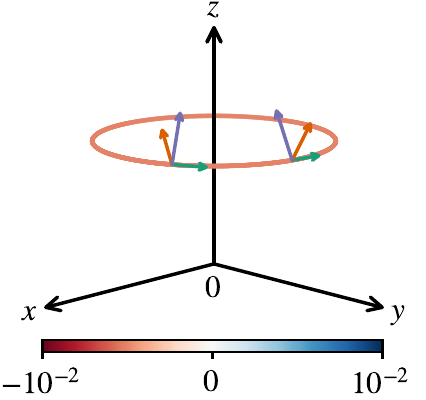}} 
\caption{For the low-dimensional nonlinear system with $\mu=0.1$, (a) details of the distance function computed on test data for the three dOTD modes; and (b) phase-space cartography of instabilities learned by the algorithm, with the color of each point on the limit cycle referring to the leading Lyapunov exponent at that point. For the dOTD modes, the same color code is used in (a) and (b): first mode, green; second mode, orange; third mode, purple.
}
\label{fig:1}
\end{figure}

Figure \ref{fig:1}a shows the distance function $d_i$ computed on part of the test set.  (Time series for the dOTD modes $\mathbf{u}_i^\textit{deep}$ and the numerically integrated OTD modes $\mathbf{u}_i^\textit{num}$ are shown in the supplementary figures S1a-c.)  The key points are that a) the neural network is able to learn the graphs $\{ \mathbf{x} \longmapsto \mathbf{u}_i(\mathbf{x})\}_{i=1}^3$ from the limited number of training points supplied to it; and b) error accumulation is inconsequential because we were careful not to terminate the optimization prematurely.  We note that $d_3$ closely follows $d_2$, a consequence of the fact that $\mathbf{u}_3^\textit{deep}$ is completely determined by $\mathbf{u}_1^\textit{deep}$ and $\mathbf{u}_2^\textit{deep}$.  We also note that Lyapunov regularization is instrumental in the optimizer converging to the stable SLB (\ref{eq:46}).  The Lyapunov exponents learned by the algorithm ($\hat{\lambda}_1= -0.005$, $\hat{\lambda}_2=-0.272$ and $\hat{\lambda}_3=-0.655$) are reasonably close to the analytical values ($\lambda_1= 0$, $\lambda_2=-0.276$ and $\lambda_3=-0.724$).

As discussed in \S\ref{sec:1}, the novelty of our approach is that the neural network provides, locally for each $\mathbf{x}$, not only the directions of strongest instabilities, but also the degree of instability along each of these directions.  This is made visually clear in figure \ref{fig:1}b, which shows all the points from the test set in the three-dimensional phase space.  Each point is colored according to its local leading Lyapunov exponent $\delta_1(\mathbf{x})=\langle \mathbf{u}_1(\mathbf{x}), \mathbf{L}(\mathbf{x}) \mathbf{u}_1(\mathbf{x}) \rangle$, as learned by the neural network.  The neural network correctly learns that $\delta_1(\mathbf{x})=0$ for all $\mathbf{x}$ on the limit cycle.  Figure \ref{fig:1}b also shows, for two of the test points, the three dOTD modes (i.e., the directions of strongest instabilities) learned by the algorithm.  Consistent with (\ref{eq:46}), the first mode is tangent to the trajectory.  Figure \ref{fig:1}b thus reinforces the claim made in \S\ref{sec:1} that our approach provides a ``cartography'' of instabilities in phase space.

\subsubsection{Charney--DeVore system}
\label{sec:42}

Next, we consider a modified version of the classical Charney--DeVore model, which describes atmospheric circulations at midlatitudes.  We consider a six-dimensional truncation of the original system, which models barotropic flow in a plane channel with topography \cite{crommelin2004mechanism}.  The governing equations are given by
\begin{subequations}
\begin{gather}
\dot{z}_1 = \gamma_1^* z_3 - C(z_1 - z_1^*) \label{cdv1} \\
\dot{z}_2 = -(\alpha_1 z_1 - \beta_1) z_3 - C z_2 - \delta_1 z_4 z_6 \\
\dot{z}_3 =  (\alpha_1 z_1 - \beta_1) z_2 - \gamma_1 z_1 - C z_3 + \delta_1 z_4 z_5 \\
\dot{z}_4 = \gamma_2^*z_6 - C(z_4 - z_4^*)  + \mu(z_2 z_6 - z_3 z_5) \\
\dot{z}_5 = -(\alpha_2 z_1 - \beta_2) z_6 - C z_5 - \delta_2 z_4 z_3 \\
\dot{z}_6 =  (\alpha_2 z_1 - \beta_2) z_2 - \gamma_2 z_4 - C z_6 + \delta_2 z_4 z_2,
\end{gather}%
\label{eq:48}%
\end{subequations}
with parameters 
\begin{subequations}
\begin{gather}
\alpha_m = 8 \sqrt{2} m^2 (b^2+m^2-1)/[\pi (4m^2-1)(b^2+m^2)], \\
\delta_m = 64 \sqrt{2} (b^2+m^2-1)/[15\pi (b^2+m^2)], \\
\mu = 16\sqrt{2}/(5 \pi), \\ 
\beta_m = \beta b^2  /(b^2+m^2), \\ 
\gamma_m = 4 \sqrt{2} m^3 \gamma b/[\pi(4m^2-1)(b^2+m^2)], \\
\gamma^*_m = 4 \sqrt{2} m \gamma b /[\pi (4 m^2-1)],
\end{gather}%
\label{eq:49}%
\end{subequations}
where $m=1$ or 2.  The parameters $\alpha_m$ and $\delta_m$ account for zonal advection in the $z_1$ and $z_4$ directions, respectively; $\beta_m$ for the so-called $\beta$ effects; $\gamma_m$ and $\gamma^*_m$ for topographic interactions; $C$ for Ekman damping; and $z_1^*$ and $z_4^*$ for zonal forcing in the  $z_1$ and $z_4$ directions, respectively.  We set $z_1^*=0.95$, $z_4^*=-0.76095$, $C=0.1$, $\beta=1.25$, $\gamma=0.2$ and $b=0.5$.  These values of the parameters give rise to significant transitions between regimes of ``zonal'' and ``blocked'' flow, resulting from nonlinear interaction between barotropic and topographic instabilities \cite{crommelin2004mechanism}.  The extreme episodes of blocked flow are of main interest to us, because they are the manifestation of transient instabilities.  Thus, it is in those intervals where we attempt to learn the OTD modes from data.

\begin{figure*}[ht!]
\centering
\subfloat[][]{\includegraphics[width=6.7in]{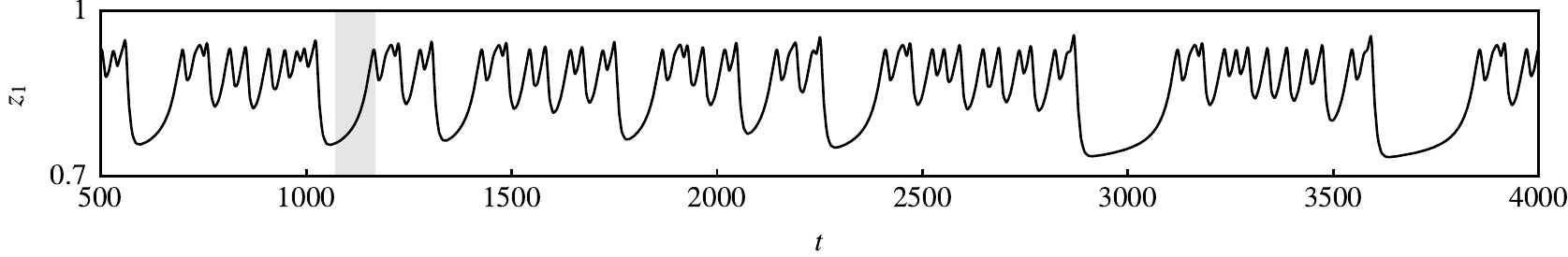}}

\subfloat[][]{\includegraphics[width=6.7in]{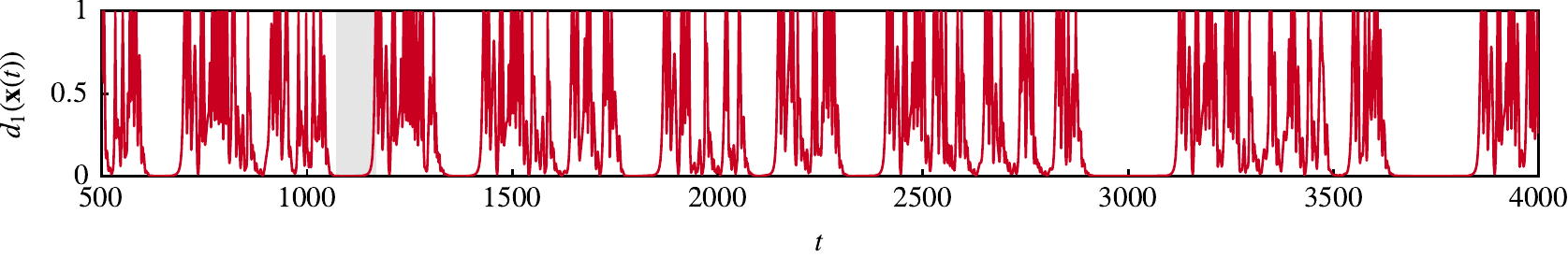}}

\caption{For the Charney--DeVore system, time series of (a) the first state coordinate $z_1$, and (b) the distance function for the first dOTD mode, with the shaded interval ($1075 \leq t \leq 1165$) identifying the training and validation sets (points outside that interval make up the test set).}
\label{fig:3}
\end{figure*}

We use a third-order Adams--Bashforth scheme with time-step size $\Delta t = 0.05$ to generate a long trajectory spanning 4000 time units.  We use zero initial conditions, except for $z_1(0)=1.14$ and $z_4(0)=-0.91$.  Snapshots are recorded at every time step, resulting in an initial dataset with 80,000 points.  For each datapoint, we reconstruct the vector field and linearized operator using its 60 nearest neighbors; then discard the first 10,000 data points corresponding to the transient regime ($0 \leq t \leq 500$).  The resulting dataset is then divided up into training, validation, and test sets. For the training set, we consider the interval $1075 \leq t \leq 1165$ during which the trajectory passes through a regime of blocked flow (see supplementary figures S2a--f), and select 50 uniformly-spaced training points in that interval.  The remaining 1750 points in that interval form the validation set.  The test set comprises all data points in the original dataset, except for those used for training and validation.  The test data therefore contains multiple episodes of blocked flow (figure \ref{fig:3}a).  We only attempt to learn the first OTD mode $\mathbf{u}_1$.  The neural network has two hidden layers, each with 128 neurons.  The learning rate for the Adam algorithm is 0.001.  Lyapunov regularization is used for the first 2000 iterations, and switched off thereafter.  Optimization is terminated at 5000 iterations.  For these parameters, the training, validation, and test errors are shown in table \ref{tab:2}.

\begin{table} 
\caption{Empirical validation results for the Charney--DeVore system and hyper-parameters given in the text.
\label{tab:2}} 
\begin{ruledtabular} 
\begin{tabular}{c|cc|cc|cc}
 	& \multicolumn{2}{c|}{Training set} 	& \multicolumn{2}{c|}{Validation set}			& \multicolumn{2}{c}{Test set}	\\  
$i$	& $\ell_i^{\textit{pde}}$	& $\bar{d}_i$	& $\ell_i^{\textit{pde}}$	& $\bar{d}_i$
	& $\ell_i^{\textit{pde}}$	& $\bar{d}_i$ \\
1	& $2.2 \cdot 10^{-5}$		& $2.0 \cdot 10^{-2}$		& $7.0 \cdot 10^{-5}$	 &	$2.5 \cdot 10^{-2}$		& $2.1 \cdot 10^{-1}$	 &	$2.6 \cdot 10^{-1}$	
\end{tabular} 
\end{ruledtabular} 
\end{table}

Figure \ref{fig:3}b shows time series for the distance $d_1(\mathbf{x}(t))$, which measures agreement between $\mathbf{u}_1^\textit{deep}$ and $\mathbf{u}_1^\textit{num}$.  Not surprisingly, the neural network is able to learn the mapping from phase space to OTD space in the interval $1075 \leq t \leq 1165$ in which the training points where supplied.  Much more remarkable is the outstanding agreement in other intervals of blocked flow (e.g., $1340 \leq t \leq 1410$ and $2900 \leq t \leq 3100$)---that is, the region in phase space synonymous with transient instabilities---showing that the neural network only needs know what a \textit{single} interval of blocked flow looks like to be able to predict all other such intervals, past or future.  This level of prediction capability is unprecedented in the context of extreme events in dynamical systems.  Figure \ref{fig:3}b and the supplementary figure S3 also show that agreement is generally poor outside intervals of blocked flow, which explains the relatively large test errors reported in table \ref{tab:2}.  This should come as no surprise, because neural networks are known to perform poorly when the testing data looks nothing like the training data.  (The fundamental assumption for generalizability is that training and testing data are drawn independently from the same probability distribution.)

Our attempts to train the neural network in an interval of zonal flow, or in an interval containing both blocked and zonal flow regimes, were unsuccessful.  We suspect that this is because the intervals of zonal flow are more ``chaotic'' (with more time scales involved) that intervals of blocked flow.  Our numerical experiments suggest that improving expressivity of the neural network (by increasing the number of hidden layers and neurons) does not solve the problem.  We note that a similar observation was made Raissi \cite{raissi2018deep}, who attempted to machine-learn the Kuramoto--Sivashinsky equation with a neural network.  Raissi \cite{raissi2018deep} noted that for this system, intervals of laminar flow posed no difficulty to the neural network, while chaotic intervals were ``stubbornly'' challenging, with the optimization algorithm not converging to the ``right values'' of the network parameters.  This description aligns with what we have seen in the present investigation of the Charney--DeVore system.  We leave investigation of this issue for future work.

\subsubsection{Flow past a cylinder}
\label{sec:43}

We conclude this section with an application of the learning algorithm to a high-dimensional dynamical system.  This is to provide an illustration of the Galerkin approach proposed in \S\ref{sec:33}.  Specifically, we consider the flow of a two-dimensional fluid of density $\rho$ and kinematic viscosity $\nu$ past a rigid circular cylinder of diameter $D$ with uniform free-stream velocity $U \mathbf{e}_x$.  The Navier--Stokes equations can be written in dimensionless form as
\begin{subequations}
\begin{gather}
\partial_t \mathbf{x} + \mathbf{x} \cdot \nabla \mathbf{x} = -\nabla p + \frac{1}{Re} \nabla^2  \mathbf{x}, \\
\nabla \cdot \mathbf{x} = 0,
\end{gather}%
\label{eq:410}%
\end{subequations}
with no-slip boundary condition ($\mathbf{x} = \mathbf{0}$) on the cylinder surface, and uniform flow ($\mathbf{x}  = \mathbf{e}_x$) in the far field.  Velocity, time and length have been scaled with cylinder diameter $D$ and free-stream velocity $U$, and the Reynolds number is $Re = U D/\nu$.  We consider the case $Re=50$, for which there exists a limit-cycle attractor which is believed to be globally and asymptotically stable \cite{noack1994global}.  Our computational approach (mesh topology, spatial discretization, and time-stepping scheme) is identical to that used by Blanchard et al. \cite{blanchard2019control,blanchard2019stabilization}.  

This flow lends itself to dimensionality reduction, because the limit-cycle attractor, while being part of an infinite-dimensional phase space, is low-dimensional, with a handful of POD modes faithfully capturing nearly all of the energy.  (In fact, the system discussed in \S\ref{sec:41} was originally introduced as a simplified model for this flow.)  Low-dimensionality of the attractor is important for leveraging the full power of the reduced-order learning algorithm proposed in \S\ref{sec:33}.  We also note that learning the dOTD modes on the limit cycle does not have much merit from the standpoint of predicting instabilities, but it does from the standpoint of flow control.  As discussed in \S\ref{sec:5}, having access to the OTD modes at any point along the periodic orbit is the stepping stone for application of the OTD control strategy proposed by Blanchard et al. \cite{blanchard2019stabilization}.

\begin{figure}[ht!]
\centering
\subfloat[][]{\includegraphics[width=3.4in]{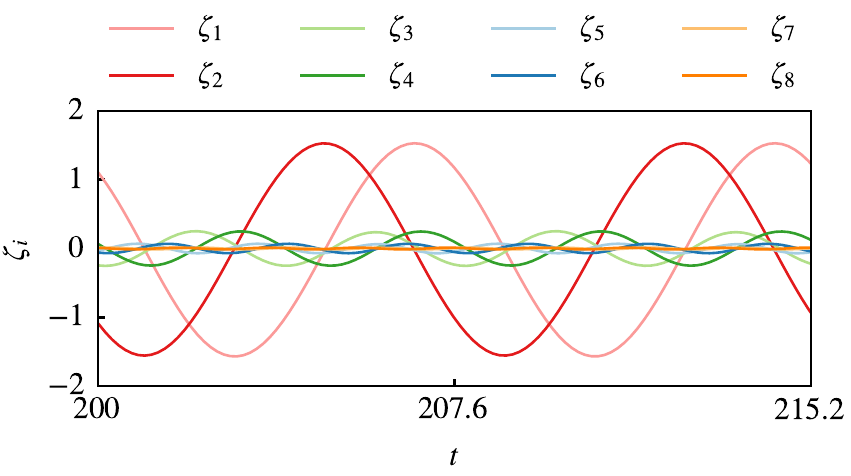}}

\subfloat[][]{\includegraphics[width=1.65in]{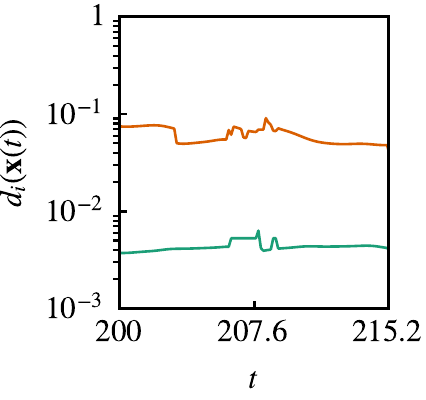}} ~
\subfloat[][]{\includegraphics[width=1.65in]{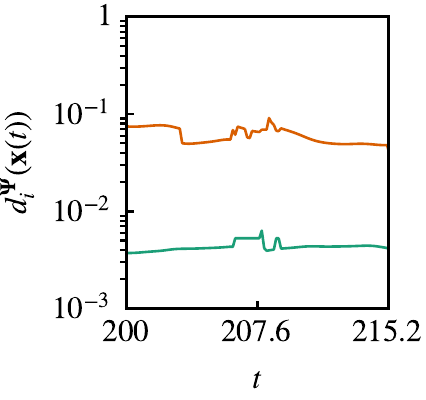}} 
\caption{For flow past a cylinder at $Re=50$, (a) details of time series for the POD coefficients; and (b) and (c) the distance functions $d_i$ and $d_i^{\bm\Psi}$, respectively, computed on test data for dOTD mode 1 (green) and 2 (orange).}
\label{fig:4}
\end{figure}

We begin with the generation of a long trajectory on the limit cycle by integrating the Navier--Stokes equations for 400 time units (corresponding to about 52 periods) with a time-step size of 0.002.  Snapshots are recorded every ten time steps, resulting in an initial dataset with 20,000 flow snapshots.  The POD modes $\bm\Phi$ are computed using 192 snapshots equally spaced over one period.  We retain the first $n_s = 8$ POD modes, accounting for more than 99.9\% of the cumulative energy.  Time series for the POD coefficients $\bm \xi \in \mathbb{R}^8$ are generated by projecting the 20,000 flow snapshots on the POD modes (figure \ref{fig:4}a).  For each point $\bm \xi_n$, we reconstruct the vector field $\mathbf{G}(\bm \xi_n)$ using an Euler-forward finite-difference approximation in $\bm\xi$-space.  (Alternatively, one can project (\ref{eq:36}) on the POD modes.)  To compute the reduced basis in which the OTD modes will be learned, we consider the local approach proposed in \S\ref{sec:33}.  We compute the tPOD modes $\{\bm\Psi(\mathbf{x}_n)\}_{n=1}^N$ using the 50 nearest neighbors of each $\mathbf{x}_n$, and then use (\ref{eq:319}) and (\ref{eq:320}) to reconstruct the reduced linearized operator and the last term on the right-hand side of (\ref{eq:317}), respectively.  We consider local tangent bases with dimension ranging from $n_t = 2$ to 6.  

Upon completion of the reconstruction step, the 20,000-point dataset in $\bm\xi$-space is divided up into training, validation, and test sets. The training set is comprised of 20 equally-spaced points over the period $50 \leq t \leq 57.6$; the validation test of 55 points over the period $100 \leq t \leq 107.6$; and the training set of 110 points over the interval $200 \leq t \leq 215.2$.  (We discard the remaining 19,815 points to avoid lengthy calculations.)  Results are presented for the first two dOTD modes.  For both, we use the same dataset splitting, and a neural network with two 32-unit hidden layers.  The Adam optimization algorithm uses a learning rate of 0.01, and is terminated after 2000 iterations.  Lyapunov regularization is turned off after 100 iterations.  For these parameters, the training, validation, and test errors are shown in table \ref{tab:3}.

\begin{table*} 
\caption{Empirical validation results for flow past a cylinder at $Re=50$ and hyper-parameters given in the text.
\label{tab:3}} 
\begin{ruledtabular} 
\begin{tabular}{c|ccc|ccc|ccc}
 	& \multicolumn{3}{c|}{Training set} 	& \multicolumn{3}{c|}{Validation set}			& \multicolumn{3}{c}{Test set}	\\ 
$i$	& $\ell_i^{\textit{pde}}$	& $\bar{d}_i$			& $\bar{d}^{\bm\Psi}_i$ 
	& $\ell_i^{\textit{pde}}$	& $\bar{d}_i$			& $\bar{d}^{\bm\Psi}_i$ 
	& $\ell_i^{\textit{pde}}$	& $\bar{d}_i$			& $\bar{d}^{\bm\Psi}_i$  \\
1	& $8.8 \cdot 10^{-5}$		& $8.5 \cdot 10^{-3}$		& $8.5 \cdot 10^{-3}$	
	& $4.3 \cdot 10^{-4}$	 	& $6.7 \cdot 10^{-3}$		& $6.7 \cdot 10^{-2}$ 
	& $6.9 \cdot 10^{-4}$		& $4.3 \cdot 10^{-3}$		& $4.3 \cdot 10^{-3}$  \\
2	& $5.7 \cdot 10^{-5}$		& $7.5 \cdot 10^{-2}$		& $7.5 \cdot 10^{-2}$		
	& $4.9 \cdot 10^{-4}$	 	& $5.5 \cdot 10^{-2}$		& $5.5 \cdot 10^{-2}$	 
	& $7.0 \cdot 10^{-4}$		& $6.0 \cdot 10^{-2}$		& $6.0 \cdot 10^{-2}$	
\end{tabular} 
\end{ruledtabular} 
\end{table*}

For $n_t=4$, figures \ref{fig:4}b,c show that the distances $\{d_i(\mathbf{x})\}_{i=1}^2$ and $\{d_i^{\bm\Psi}(\mathbf{x})\}_{i=1}^2$ are virtually zero at the test points.  This shows that a) the error introduced by the low-dimensional reconstruction of the tangent space is negligible, and b) the neural network finds the best representation of the OTD modes in that reduced tangent space.  Supplementary figures S4a--f provide visual confirmation that the dOTD modes learned in the reconstructed tangent space are indistinguishable from their numerically integrated counterparts.  These results illustrate the benefits of learning the OTD modes in a reduced subspace.  Equally good agreement was obtained for $n_t=2$ and 6 with the same network parameters.

\section{Discussion}
\label{sec:5}

We now discuss possible improvements and modifications to the OTD learning algorithm introduced in \S\ref{sec:3}, as well as implications for data-driven control of instabilities in dynamical systems.

In the examples discussed in \S\ref{sec:4}, the nonlinearity appearing in the governing equations was no stronger than quadratic with respect to the state variables, and the neural network was able to learn the OTD graph using the state $\mathbf{x}$ as its only ``active'' input.  In cases in which the nonlinearity is known to be, or suspected to be, stronger than quadratic (e.g., with terms involving higher-order polynomials, trigonometric functions, or nonlinear differential operators), it is likely that supplying $\mathbf{x}$ as the only input will call for wider, deeper networks than used in \S\ref{sec:4}.  To keep the number of network parameters reasonably small and facilitate convergence of the optimization algorithm, one possibility is to use as additional inputs a library of nonlinear functions of the state; for example, $\{\mathbf{x}_n, \mathbf{F}(\mathbf{x}_n),  \sin(\mathbf{x}_n), \exp(-\mathbf{x}_n^2), \mathbf{x}_n \! \cdot \!\nabla \mathbf{x}_n\}_{n=1}^N$.  This approach is in the same spirit as the SINDy algorithm \cite{brunton2016discovering}, in which sparse regression is applied to a library of nonlinear functions of $\mathbf{x}$ in order to discover governing equations from state measurements.  We note that equation discovery goes well beyond the SINDy algorithm, with more general techniques such as grammar-based equation discovery \cite{todorovski1997declarative} and process-based modeling \cite{todorovski2006integrating,tanevski2012biocircuit}.

We also note that in any laboratory experiment, sensing capabilities are limited by the apparatus, leading to errors in state estimation and reconstruction.  Uncertainty in state measurements may be accounted for by trading the neural-network approach for one based on Gaussian processes (GPs), because GPs have the advantage of providing error estimates at each testing points.  GPs have been found capable of handling sizable noise levels in a number of problems similar to the present, including deep learning of partial differential equations and discovery of governing equations from noisy measurements \cite{raissi2017inferring,raissi2018hidden}.  Another possibility is to use an approach based on reservoir computing, a type of neural-network architecture on which noise in the dataset has, paradoxically, a stabilizing effect.  Vlachas et al. \cite{vlachas2020recurrent} recently applied reservoir computing to the problem of predicting chaotic dynamics, and found that addition of noise in the training data not only led to better generalization, but also improved performance of the network on both the training and testing datasets.

The method proposed in \S\ref{sec:3} is fully data-driven, in the sense that no input other than state snapshots is required to learn the dOTD modes.  (The vector field and linearized operator are reconstructed using nothing but state snapshots.)  If governing equations \textit{are} available (either derived from first principles or reconstructed from data), then there is another possibility to learn the graphs $\mathbf{x} \longmapsto \mathbf{u}_i(\mathbf{x})$; that is, generate a large number of $\{\mathbf{x}_n, \mathbf{u}_i(\mathbf{x}_n)\}$ pairs by solving the state and OTD equations numerically, and model the input--output relationship by a neural network, whose parameters are found by minimizing the empirical risk
\begin{equation}
\ell_i^\textit{emp}(\bm\theta_i) = \frac{1}{N} \sum_{n=1}^N \ell(\mathbf{x}_n, \mathbf{u}_i(\mathbf{x}_n;\bm\theta_i)),
\end{equation}
where $\ell$ is an appropriate loss function (e.g., quadratic loss or log-cosh loss).  To $\ell_i^\textit{emp}$ may be appended the loss function (\ref{eq:31}), which  then acts as a physics-informed \cite{raissi2018hidden,zhang2019learning} (equation-based) regularization term.  The downsides of this approach are that a) access to governing equations is mandatory, and b) generating the input--output pairs requires solving $r+1$ $d$-dimensional initial-boundary-value problems, which may be computationally expensive.

From the standpoint of predicting instabilities, the benefit of learning the dOTD modes from data is that it gives access to directions of instabilities at any point in phase space, and to the leading Lyapunov exponents, regardless of their sign.  But the dOTD learning algorithm also has significant implications from the standpoint of controlling instabilities.  As discussed in \S\ref{sec:21}, we have recently shown that the OTD modes can be incorporated into reduced-order control algorithms for stabilization of unsteady high-dimensional flows.  The OTD control strategy proposed in Ref. \onlinecite{blanchard2019stabilization} requires solving the OTD equations concurrently with the state equations, because the control force belongs to the OTD subspace.  With the dOTD learning approach, this requirement disappears, because the neural-network approach can be used to build a library of dOTD modes for various regions of the phase space.  (Construction of the OTD library may be done offline.)  Then, as the controlled trajectory wanders about in phase space, the controller can look up in the OTD library the dOTD modes associated with the current state.  (If the trajectory visits a state that is not present in the library, then one can interpolate between nearby states for which dOTD modes are available.)  Library look-up can be done in real time because the computational complexity of the look-up algorithm scales with the dimension of the OTD subspace, which makes the approach very attractive from the standpoint of controlling high-dimensional systems in real time.  We note that similar ideas were employed in the context of nonlinear model order reduction by Amsallem et al. \cite{amsallem2008interpolation, amsallem2012nonlinear}.

Finally, there is the issue of how the predictive capabilities of the neural network are affected by changes in system parameters.  We first note that even a small change in system parameters has the potential of considerably altering the topology of the phase space, and in particular, the number, nature, and properties of the attractors.  (This is apparent when a bifurcation occurs.)  We also note that one of the prerequisites for the neural network to perform well on unseen data is that the training and testing data be drawn from the same underlying probability distribution.  Large variations in system parameters are likely to violate this assumption, seriously compromising generalizability of the neural network.  Small variations in system parameters may lead to good generalizability provided that the changes in phase-space topology and underlying probability distribution are also small.

\section{Conclusion}
\label{sec:6}

For a large class of dynamical systems, the optimally time-dependent (OTD) modes, a set of deformable orthonormal tangent vectors that track directions of instabilities along any trajectory, are known to depend pointwise on the state of the system on the attractor, and not on the history of the trajectory.  We have developed a learning algorithm based on neural networks to learn this pointwise mapping from phase space to OTD space using data collected along one or more trajectories of the system.  The proposed method is fully data-driven as it requires no other input than snapshots of the state, and is applicable regardless of the dimensionality of the system.  The learning process provides a cartography of directions associated with strongest instabilities in phase space, as well as accurate estimates for the leading Lyapunov exponents of the attractor.  This has significant implications for data-driven prediction of dynamical instabilities, with the learning algorithm exhibiting predictive capabilities of extreme events to a degree that is unprecedented, but also for design and implementation of reduced-order controllers capable of operating in real time.

\section*{Supplementary material}
See Supplementary Material for supplementary figures referenced
in the paper.

\section*{Acknowledgments}
The authors acknowledge discussions with Dr. Hassan Arbabi.  This work was supported by the Army Research Office (Grant No. W911NF-17-1-0306).

\section*{References}
\bibliography{bibl}

\end{document}